\documentclass[12pt]{iopart}
\expandafter\let\csname equation*\endcsname=\relax 
\expandafter\let\csname endequation*\endcsname=\relax 
\usepackage{amsmath,amssymb,amsthm,bm,braket,txfonts,graphicx}
\usepackage[square,numbers,sort&compress]{natbib}

\usepackage[bookmarks=true,colorlinks,citecolor=blue,urlcolor=blue]{hyperref}
\usepackage{colortbl}
\definecolor{gray}{RGB}{230,230,230}

\begin{document}

\title[Classification of magnon thermal Hall systems]{Classification of magnon thermal Hall systems based on U(1) to non-Abelian gauge fields}
\author{Masataka Kawano and Chisa Hotta}
\address{Department of Basic Science, University of Tokyo, Meguro-ku, Tokyo 153-8902, Japan}
\ead{masatakakawano@g.ecc.u-tokyo.ac.jp}
\vspace{10pt}
\begin{indented}
\item[]\today
\end{indented}

\begin{abstract}
Magnon thermal Hall effect in insulating magnets is the manifestation of Berry curvature in magnon bands, 
which is formulated using the emergent gauge fields that act on magnons as a fictitious magnetic field. 
In ferromagnets, it is commonly accepted as the outcome of U(1) gauge fields generated 
by Dzyaloshinskii-Moriya interactions and spin textures, 
but this mechanism is often suppressed by symmetry-enforced cancellations in many lattice geometries, 
known as a no-go rule. 
As a result, antiferromagnetic insulators have long been considered as unfavorable platforms for the effect. 
We show that antiferromagnets with multiple magnetic sublattices
naturally host non-Abelian SU($N$) gauge fields in magnon band structures,
providing a robust rule-to-go mechanism. 
The noncommutativity of these gauge fields prevents Berry-curvature cancellation
and guarantees a nonvanishing thermal Hall response. 
As a minimal realization, we demonstrate that a coplanar $120^\circ$
antiferromagnet with Dzyaloshinskii--Moriya interactions
constitutes a canonical SU(3) platform for the magnon thermal Hall effect.
We provide a table of so-far-known two-dimensional lattice geometries and variants of magnetic structures, 
along with the corresponding gauge fields, 
providing a unified guideline for identifying magnetic materials, 
including antiferromagnets and altermagnets, that host thermal Hall transport. 
\end{abstract}
\vspace{2pc}
Keywords: magnons, thermal Hall effect, emergent gauge fields
\maketitle

\section{Introduction}
The magnon thermal Hall effect was first experimentally observed around 2010 
in the pyrochlore ferromagnet Lu$_2$V$_2$O$_7$~\cite{onose2010science}, 
alongside the celebrated formula of the thermal Hall conductivity of magnons, 
$\kappa_{xy}$, derived by Matsumoto and Murakami~\cite{matsumoto2011prl,matsumoto2011prb}, 
which revealed a transparent physical structure despite the underlying 
technical complexity of its derivation: 
$\kappa_{xy}$ is expressed as a Brillouin-zone integral of the Berry curvature of magnon bands
weighted by functions analogous to their Bose distribution functions.
This compact form, closely analogous to the electronic Hall conductivity $\sigma_{xy}$, 
provided a powerful framework for identifying microscopic mechanisms that 
generate Berry curvature in magnon band structures.

Early studies focused primarily on ferromagnetic insulators,
where the antisymmetric Dzyaloshinskii-Moriya (DM) interaction and magnetically ordered spin textures
introduce complex Peierls phases into magnon hopping amplitudes~\cite{onose2010science}.
These phases can be interpreted as arising from an effective U(1) gauge field,
corresponding to a fictitious magnetic flux experienced by magnons traversing closed loops on the lattice.
However, for edge-sharing lattice geometries,
the U(1) gauge flux typically appears with opposite signs on symmetry-related loops,
or equivalently as Berry curvatures of opposite sign in different regions in momentum space. 
As a consequence, their contributions cancel exactly, leading to a vanishing thermal Hall conductivity~\cite{katsura2010prl,ideue2012prb}.
This symmetry-enforced cancellation is commonly referred to as a no-go rule for the U(1) gauge-field mechanism. 
Material realizations and model studies were therefore concentrated on magnets 
on corner-sharing lattices such as kagome~\cite{katsura2010prl,hirschberger2015prl,mook2014prb1,mook2014prb2,mook2016prb,owerre2017prb,owerre2018prb,seshadri2018prb,laurell2018prb,owerre2017epl,owerre2018sp,owerre2019ap}, pyrochlore~\cite{onose2010science,ideue2012prb,laurell2017prl}, or distorted triangular lattices~\cite{kim2019prb,kim2024ncom}. 

A further fundamental difficulty arises from the fact that most Mott insulating magnets are antiferromagnets,
stabilized by dominant kinetic exchange processes that favor antiferromagnetic Heisenberg couplings.
A similar issue was long recognized in electronic systems:
it was traditionally believed that a finite magnetization $\bm{M}\neq 0$, as realized in ferromagnets,
is required to produce a nonvanishing anomalous Hall effect (AHE)~\cite{nagaosa2010rmp}. 
In electronic systems, spin-orbit coupling (SOC) generates a Berry curvature $\Omega(\bm{k})$
in energy bands, but a breaking of time-reversal symmetry (TRS) is essential. 
If TRS is present, $\Omega(\bm{k})=-\Omega(-\bm{k})$, 
leading to an exact cancellation in $\sigma_{xy}$. 
Antiferromagnets, despite having zero net magnetization,
often possess a combined symmetry of TRS and lattice translation (or other spatial operations),
which acts as an effective or pseudo TRS.
This symmetry enforces a strict no-go rule for the AHE in conventional antiferromagnets.
The first proposal circumventing the no-go rule was achieved in 2014 
for noncollinear antiferromagnets~\cite{hua2014prl}. 
Even with $\bm{M}=0$, the AHE can arise provided that 
the magnetic symmetry of the system is low enough to break the pseudo TRS. 
This insight was subsequently confirmed experimentally in noncollinear antiferromagnets 
such as Mn$_3$Sn and Mn$_3$Ge~\cite{nakatsuji2015nature,nayak2016sciad,kiyohara2016prap,liu2017srep}, 
establishing that Berry curvature and Hall responses are governed by symmetry, rather than by net magnetization alone. 

The extension of this finding to magnon thermal Hall effects in antiferromagnetic insulators,
however, required a longer timespan. 
This is because, although some studies phenomenologically referred to the analogy of 
noncollinear insulating kagome magnets with these electronic systems~\cite{owerre2017prb,owerre2017epl,owerre2018prb,laurell2018prb}, 
making formulative correspondence between electronic AHE and magnon thermal Hall effects was not straightforward. 
A canonical theoretical platform was first proposed in 2019 by the authors:
a square-lattice antiferromagnet exhibiting a finite magnon thermal Hall effect~\cite{kawano2019prb1},
motivated by the experimental observation of microwave nonreciprocity
in the related material Ba$_2$MnGeO$_7$~\cite{iguchi2018prb}.
The key insight is the correspondence between two-sublattice antiferromagnets
and Rashba-Dresselhaus electrons.
Within linear spin-wave theory (LSW), spin fluctuations on the two sublattices form distinct magnon species,
which can be interpreted as pseudospin degrees of freedom.
DM interactions between sublattices then act as a pseudo SOC,
generating band splitting, momentum-space spin textures~\cite{kawano2019comphys,kawano2019prb2},
and a finite thermal Hall response~\cite{kawano2019prb1}.
Formally, such spin-mixing hoppings are described not by a U(1) gauge field,
but by an SU(2) gauge field represented by $2\times2$ spin-rotation matrices,
closely paralleling the electronic SOC framework~\cite{frohlich1993rmp}.
From an application perspective, magnon Hall transport is particularly attractive for spintronics,
as magnons can carry heat and angular momentum without Joule heating,
offering promising routes toward low-dissipation information transport~\cite{chumak2015natphys}.

It was only very recently that a genuine experimental observation of the magnon thermal Hall effect
in an insulating antiferromagnet was reported, namely in MnSc$_2$S$_4$ in an applied magnetic field~\cite{takeda2024ncom},
within a phase exhibiting a three-sublattice antiferromagnetic skyrmion texture~\cite{comment-triangular}.
We have attributed the observed response to an emergent SU(3) gauge field
acting on three magnon species, providing a similar picture as the SU(2) gauge field for the earlier theory on square-lattice antiferromagnets.
Still, such examples are rare, reflecting the difficulty of identifying materials
with appropriate energy scales and magnetic textures that naturally realize higher-rank gauge structures. 
In that respect, our previous theories relied on model-specific features, 
and a unified guiding principle for material exploration has been lacking. 

In this paper, we address this gap by providing a comprehensive classification of 
two-dimensional (2D) magnetic insulators realized on representative edge-sharing and corner-sharing lattices.
We summarize the conditions under which a nonvanishing thermal Hall effect is expected, 
and identify its microscopic origin in terms of emergent gauge fields. 
In Sec.~\ref{sec:u1}, we review the U(1) gauge-field mechanism 
and clarify how symmetry leads to the no-go rule in many ferromagnetic systems.
In Sec.~\ref{sec:nonabelian}, we develop the general framework for higher-rank non-Abelian gauge fields
in sublattice-based antiferromagnets,
and show that the noncommutativity of these gauge fields
provides a natural and robust route to circumvent the no-go rule.
As a concrete demonstration, 
we present the thermal Hall effect in a canonical antiferromagnet
with coplanar $120^\circ$ magnetic order,
illustrating the essential physics of this higher-rank gauge-field mechanism.

\begin{table}[t]
    \centering
    \footnotesize{
    \begin{tabular}{lllllll} \br
    Lattice & Order & Gauge field & Origin & $\kappa_{xy}^{(\mathrm{theory})}$ & $\kappa_{xy}^{(\mathrm{exp.})}$ & Ref\\ \mr
    Square & FM & U(1) & DM & -- & -- & \cite{kawano2019prb1,ideue2012prb}\\
    Square & FM & U(1) [${\bar{\phi}\neq0}$] & AC & $\checkmark$ & & \cite{nakata2017prb1}\\
    Square & CL-AFM & U(1) & DM & -- & & \cite{kawano2019prb1}\\
    Square & CL-AFM & U(1) & SSC & -- & & \cite{katsura2010prl}\\
    Square & CL-AFM & U(1) [${\bar{\phi}\neq0}$] & AC & $\checkmark$ & & \cite{nakata2017prb2}\\
    \rowcolor{gray} Square & NCL-AFM & SU(2) & DM & $\checkmark$ & & \cite{kawano2019prb1}\\
    Triangular & FM & U(1) & SSC & -- & & \cite{katsura2010prl} \\
    Triangular & NCL-AFM & U(1) & SSC & -- & & \cite{katsura2010prl} \\
    \rowcolor{gray} Triangular & NCL-AFM & SU(3) & DM & $\checkmark$ & & This study\\
    Triangular & FM-SkX & U(1) [${\bar{\phi}\neq0}$] & FM-SkX & $\checkmark$ & $\checkmark$ & \cite{hoogdalem2013prb,kong2013prl,iwasaki2014prb,oh2015prb,roldan2016njp,mook2017prb,kim2019prl,nikolic2020prb,akazawa2022prr} \\
    \rowcolor{gray} Triangular & AFM-SkX & SU(3) & AFM-SkX & $\checkmark$ & $\checkmark$ & \cite{takeda2024ncom}\\
    Triangular (distorted) & NCL-AFM & U(1) & SSC & $\checkmark$ & & \cite{kim2019prb,kim2024ncom}\\
    Kagome & FM & U(1) & DM & $\checkmark$ & $\checkmark$ & \cite{mook2014prb1,mook2014prb2,hirschberger2015prl,seshadri2018prb,owerre2018sp} \\
    Kagome & FM & U(1) & SSC & $\checkmark$ & & \cite{katsura2010prl,mook2014prb1,mook2014prb2,mook2016prb,seshadri2018prb,owerre2018sp} \\
    Kagome & NCL-AFM & U(1) & AC & $\checkmark$ & & \cite{owerre2019ap} \\
    Kagome & NCP-AFM & U(1) & DM & $\checkmark$ & & \cite{laurell2018prb} \\
    Kagome & NCP-AFM & U(1) & SSC & $\checkmark$ & & \cite{owerre2017prb,owerre2017epl,owerre2018prb,laurell2018prb} \\
    Pyrochlore & FM & U(1) & DM & $\checkmark$ & $\checkmark$ & \cite{onose2010science,ideue2012prb} \\
    Pyrochlore & NCP-AFM & U(1) & DM+SSC & $\checkmark$ & $\checkmark$ & \cite{laurell2017prl} \\
    Honeycomb & FM & U(1) & DM & $\checkmark$ & & \cite{owerre2016jap,owerre2016prb,lu2021prl} \\
    Honeycomb & FM & U(1) & Kitaev/Gamma & $\checkmark$ & & \cite{mcclarty2018prb,zhang2021prb,chern2021prl}\\
    Honeycomb & CL-AFM & U(1) & Kitaev/Gamma & $\checkmark$ & & \cite{neumann2022prl}\\
    Honeycomb & NCL-AFM & U(1) & DM & $\checkmark$ & & \cite{owerre2017jap}\\
    Honeycomb & NCP-AFM & U(1) & SSC & $\checkmark$ & & \cite{owerre2017jpcm,chern2020prr,chern2021nqm,koyama2021prb} \\
    Lieb & FM & U(1) & DM & $\checkmark$ & & \cite{cao2015jpcm}\\
    Star & NCP-AFM & U(1) & SSC & $\checkmark$ & & \cite{owerre2016jpcm}\\
    Dice & FM & U(1) & DM/PD & $\checkmark$ & & \cite{debnath2025prb1}\\
    \rowcolor{gray} Shastry-Sutherland & VBC & SU(2) & DM & $\checkmark$ & -- & \cite{judit2015ncom,malki2019prb,suetsugu2022prb}\\
    \rowcolor{gray} Square-dimer & VBC & SU(2) & DM & $\checkmark$ & & \cite{buzo2024prb}\\
    \rowcolor{gray} Checkerboard & CL-AFM (alter) & SU(2) & DM & $\checkmark$ & & \cite{hoyer2025prb}\\ \br
    \end{tabular}
    \caption{Classification of magnon thermal Hall systems discussed in the literature.
    $\kappa_{xy}^{(\mathrm{theory})}$ and $\kappa_{xy}^{(\mathrm{exp.})}$ denote whether the magnon thermal Hall effect is investigated theoretically or observed experimentally. $\checkmark$ (--) indicates the presence (absence) of the thermal Hall conductivity. Abbreviations: FM (ferromagnetic), AFM (antiferromagnetic), CL (collinear), NCL (noncollinear but coplanar), NCP (noncoplanar), SkX (skyrmion crystal), VBC (valence-bond crystal), DM (Dzyaloshinskii-Moriya), SSC (scalar spin chirality), PD (pseudo-dipolar), and AC (Aharonov-Casher). U(1) [${\bar{\phi}\neq0}$] indicates the presence of a uniform flux. Shaded rows highlight systems where non-Abelian gauge fields (SU(2) or SU(3)) enable a finite thermal Hall effect on edge-shared lattices.}
    \label{table:classification}
    }
\end{table}

\section{Thermal Hall classification of lattice models}
\subsection{Preliminaries}
We begin by briefly outlining the theoretical formulation of the magnon thermal Hall effect.
We consider localized spin moments 
$\hat{\bm{S}}_{i}=(\hat{S}_{i}^{x},\hat{S}_{i}^{y},\hat{S}_{i}^{z})$
on lattice sites, interacting with neighboring spins.
The minimal quantum spin Hamiltonians known to host a thermal Hall response
are typically composed of the isotropic Heisenberg exchange
and bond-antisymmetric DM interactions,
\begin{align}
    \hat{\mathcal{H}}
    =\sum_{\braket{i,j}}\left[
    J\,\hat{\bm{S}}_{i}\cdot\hat{\bm{S}}_{j}
    +\bm{D}_{i,j}\cdot(\hat{\bm{S}}_{i}\times\hat{\bm{S}}_{j})
    \right]
    +\cdots ,
    \label{eq:H-AFM-chain}
\end{align}
where $\braket{i,j}$ is the pair of nearest-neighbor $i$ and $j$ sites and the ellipsis denotes additional interactions such as 
single-ion anisotropy or bond-symmetric anisotropic exchanges
including Kitaev and $\Gamma$ terms. 
While these additional couplings are often subleading, they play an essential role in stabilizing magnetic order,
whose symmetry and texture ultimately determine whether a thermal Hall effect can appear. 
\par
Our analysis proceeds along different layers of conceptual levels.
First, for a given spin Hamiltonian, we restrict ourselves to magnetically ordered ground states
and focus on the lowest-order fluctuations around them. 
These excitations are described either by standard LSW
or, equivalently, by an effective low-energy field theory at the noninteracting level.
The LSW formulation is well established and reviewed in standard textbooks,
while the corresponding field-theoretical framework 
is summarized in Sec.~\ref{sec:nonabelian} and discussed in more detail in Refs.~\cite{kawano2025prb,kawano2025axv}.

\par
Accordingly, the starting point of our analysis is a magnetically ordered ground state
amenable to an LSW treatment. 
At present, only a limited number of analytical frameworks exist 
to describe transport phenomena in magnetically disordered or spin-liquid phases,
where excitations typically form a continuum of fractionalized spinons~\cite{zhang2024pr}.
In contrast, for ordered magnets, fluctuations of the ordered moments can be systematically treated
by applying the Holstein-Primakoff (H-P) transformation~\cite{holstein1940pr},
\begin{align}
    \hat{\bm{S}}_{i}\simeq\sqrt{S}(\hat{b}_{i}\bm{e}_{i}^{-}+\mathrm{H.c.})+(S-\hat{b}_{i}^{\dagger}\hat{b}_{i})\bm{e}_{i}^{Z},
    \label{eq:HPtr}
\end{align}
where $S$ is the spin quantum number, $\hat{b}_{i}$ ($\hat{b}_{i}^{\dagger}$) 
is the annihilation (creation) operator of magnons, 
and $\bm{e}_{i}^{\pm}=(\bm{e}_{i}^{X}\pm\mathrm{i}\bm{e}_{i}^{Y})/\sqrt{2}$ with 
$\bm{e}_{i}^{X/Y/Z}$ being the local orthogonal three-dimensional (3D) unit vectors \textit{in spin space}.
Namely, we assume that the ordered spin moment points in the $Z$-direction. 
For collinear magnetic orders, the $Z$ axis is common to all sites and coincides with a global spin quantization axis.
For noncollinear or noncoplanar spin textures, however,
the local $Z$ axis depends on the site, reflecting the spatial variation of the ordered moments.
\par
Diagonalization of the resulting quadratic magnon Hamiltonian yields a set of magnon bands. 
The Berry curvature on the $n$th magnon band, $\Omega_{xy}^{(n)}(\bm{k})$, is finite if there exists an imaginary component 
of the eigenstates, and the thermal Hall conductivity in 2D systems takes the form~\cite{matsumoto2011prb,matsumoto2011prl,matsumoto2014prb}, 
\begin{align}
    \kappa_{xy}=-\frac{k_{B}^{2}T}{\hbar}\int_{\mathrm{BZ}}\frac{d^{2}\bm{k}}{(2\pi)^{2}}
    \sum_{n}\left\{c_{2}[f_{\mathrm{B}}(\varepsilon_{n}(\bm{k}))]-\frac{\pi^{2}}{3}\right\}\Omega_{xy}^{(n)}(\bm{k}), 
    \label{eq:Thermal-Hall}
\end{align}
where $f_{\mathrm{B}}(\varepsilon)=1/[\mathrm{exp}(\varepsilon/k_{B}T)-1]$ is the Bose distribution function
and $c_{2}[x]=\int_{0}^{x}\ dt[\mathrm{ln}\{(1+t)/t\}]^{2}$ is the monotonically decreasing function. 
This expression closely parallels that of the electronic anomalous Hall conductivity,
$\sigma_{xy}=(-e^{2}/\hbar)\int_{\mathrm{BZ}}[\mathrm{d}^{2}\bm{k}/(2\pi)^{2}]\ \Omega(\bm{k})f_{\mathrm{F}}(\varepsilon_{n}(\bm{k}))$,
where the Berry curvature is integrated over the Brillouin zone
with the Fermi distribution function $f_{\mathrm{F}}(\varepsilon)=1/[\mathrm{exp}(\varepsilon/k_{B}T)+1]$. 
\par
Despite this formal analogy, there exists a fundamental difference between
the magnon thermal Hall conductivity $\kappa_{xy}$ and its electronic counterpart $\sigma_{xy}$.
In fermionic systems, the Hall response is governed primarily by states near the Fermi level,
which can lead to quantized values of $\sigma_{xy}$ in topologically nontrivial phases~\cite{nagaosa2010rmp}. 
In contrast, magnons are bosonic excitations, 
and their thermal occupation extends over a broad energy range at finite temperature.
As a result, Berry curvatures of low to high energy bands can simultaneously contribute significantly to $\kappa_{xy}$,
and quantization is absent, except in very special cases where the low-energy excitations
admit a fermionic description, such as in Majorana representations~\cite{kitaev2006ap}.

\subsection{Classification of systems with thermal Hall effect}
Our major goal is to clarify when the momentum-space integration in Eq.~\eqref{eq:Thermal-Hall} yields a nonvanishing result.
This depends both on the mechanism generating sufficiently large Berry curvature
and on the avoidance of symmetry-enforced cancellations. 
In Table~\ref{table:classification}, we summarize representative lattice geometries and magnetic orders,
and classify whether they exhibit $\kappa_{xy}\neq0$ or $\kappa_{xy}=0$.

For square-lattice systems, both ferromagnets and antiferromagnets generally exhibit $\kappa_{xy}=0$
when the magnetic order is collinear,
reflecting a no-go rule analogous to that for the electronic AHE.
Noncollinear antiferromagnets can circumvent this constraint, but the underlying mechanism relies on SU(2) gauge fields
rather than a U(1) description.
External manipulation, such as through the Aharonov-Casher (AC) effect, can further lower the symmetry and provide exceptional routes
to finite thermal Hall responses~\cite{nakata2017prb1,nakata2017prb2}. 

On the triangular lattice, the no-go constraint is even more restrictive, as ferromagnetic and noncollinear antiferromagnetic orders
still preserve a $\pi$-rotation symmetry that enforces a pseudo TRS.
In contrast, skyrmion crystals remain favorable platforms for $\kappa_{xy}\neq0$ due to their topologically-nontrivial spin textures. 
The recently discovered antiferromagnetic skyrmion crystal (AFM-SkX) phase in MnSc$_{2}$S$_{4}$ represents an especially rich example~\cite{reil2002jac,fritsch2004prl,gao2017nphys,gao2020nature}, 
combining three magnetic sublattices and the spin textures, which cooperatively generate an SU(3) gauge. 

Kagome and pyrochlore lattices, by contrast, readily support finite $\kappa_{xy}$ within a U(1) gauge-field framework,
as the symmetry conditions underlying the no-go rule do not apply. 
Although the honeycomb lattice has an edge-sharing geometry, it is exceptional in that next-nearest-neighbor interactions,
familiar from the Kane-Mele model~\cite{kane2005prl1,kane2005prl2}, naturally yield the U(1) flux pattern that allows for thermal Hall responses.

More complex lattices, including Lieb and Shastry-Sutherland lattices, can in principle also host a thermal Hall effect. 
However, despite theoretical predictions~\cite{judit2015ncom}, experimental realizations such as SrCuBO$_{4}$
exhibit vanishing or extremely small $\kappa_{xy}$ within current experimental resolution~\cite{suetsugu2022prb}.
While this discrepancy highlights the limitations of idealized models,
the classification summarized in Table~\ref{table:classification} provides a practical guideline for identifying material platforms.

Recently, another class of magnetic materials called altermagnets has been focused~\cite{hayami2020prb,smejkal2022prx1,smejkal2022prx2}. 
Altermagnets are two-sublattice collinear antiferromagnets that lack symmetries relating the two magnetic sublattices,
leading to giant momentum-dependent spin splitting in electronic band structures even in the absence of SOC.
While this concept was originally introduced in the context of electronic systems, 
it is naturally extended to magnon excitations. 
Indeed, the collinear antiferromagnet on a checkerboard lattice provides a minimal realization of an insulating altermagnet.
The inequivalence of the two magnetic sublattices prohibits the symmetry operations 
that enforce $\kappa_{xy}=0$, allowing for finite $\kappa_{xy}$~\cite{hoyer2025prb}. 
In that respect, the altermagnets only differ from the antiferromagnets in terms of symmetry operation, 
and are included in the SU(2) framework we present in \S.\ref{sec:nonabelian}. 

In the following two sections, we systematically discuss the microscopic origins of
U(1) and SU($N$) gauge fields in magnon systems, and clarify the conditions under which
a finite thermal Hall conductivity can be expected.

\section{Effective U(1) gauge-field}
\label{sec:u1}
\begin{figure}[t]
    \centering
    \includegraphics[width=155mm]{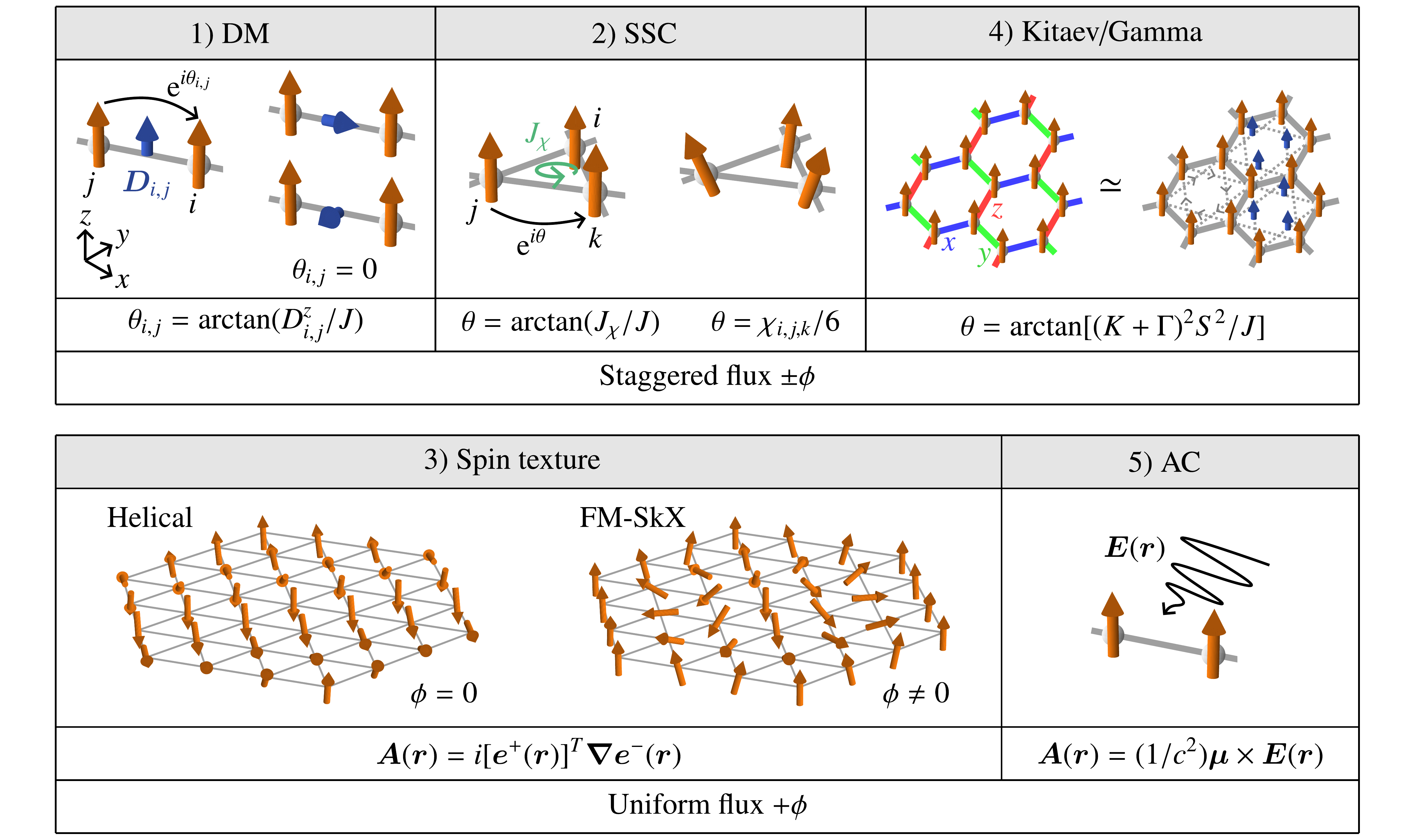}
    \caption{Origins of the U(1) gauge field: Dzyaloshinskii-Moriya (DM) interaction, scalar-spin chirality (SSC), Kitaev and Gamma interactions, spin texture, and Aharonov-Casher (AC) effect. The DM interaction, SSC, and Kitaev and Gamma interactions generate staggered flux pattern, whereas the spin texture and AC effect can generate uniform flux pattern.}
    \label{fig:u1-gauge}
\end{figure}

A series of theoretical works ascribing the thermal Hall effect to the U(1) gauge was done around 2010~\cite{fujimoto2009prl,katsura2010prl,matsumoto2011prl,matsumoto2011prb} coherently with the first experimental observation of a magnon Hall effect~\cite{onose2010science}. 
The physical origins of the U(1) gauge field known so far are listed as 
\textit{1) DM interaction, 2) $\&$ 3) noncoplanar spin textures including helical and skyrmionic states, 4) anisotropic exchange interactions 
such as Kitaev and $\Gamma$ terms, and 5) the couplings with the electric field} (see Fig.~\ref{fig:u1-gauge}). 
The emergent U(1) gauge field acts on the magnons in real space and endows a finite Berry curvature of their energy bands in reciprocal space.
However, depending on the symmetry of the system, the Berry curvature often suffers a cancellation and 
zeros out their contributions to the thermal Hall effect. 
\par
Notice that we mean by this symmetry not necessarily the one coming from the original crystal symmetry; 
quite often, there are cases where the thermal Hall conductivity vanishes within the framework of LSW, 
even though the original spin Hamiltonian seemingly allows for a finite value from a symmetry viewpoint. 
This is because, part of the pre-broken symmetry of the spin Hamiltonian is restored at the lowest order approximation, 
in particular, at the noninteracting level of the LSW. 
In such a case, the inclusion of higher-order terms yielding magnon-magnon interactions generally puts the system back to the original symmetry, 
which is likely the case at a sufficiently high temperatures (or for cases including higher excited states)~\cite{zhitomirsky2013rmp}. 
Whilst, such an effect, at the crudest level, would be quantitatively small by orders of magnitude 
and does not change the observation much except for a few particular cases. 
How the restored symmetry is put back to the original symmetry by the inclusion of higher-order interactions 
and affects the phenomena have been discussed, for example in Refs.~\cite{mook2021prx,chatzichrysafis2025prb}.
Notice that the symmetry-restoration here means the symmetry after the magnetic ordering takes place in the original Hamiltonian. 
Indeed for some cases like antiferromagnetic ordering we discuss in Sec.~\ref{sec:nonabelian}, 
the translational symmetry as well as other symmetries are broken at the starting point of the LSW. 
\par
In the following, we focus on the lowest-order noninteracting level, 
first summarizing how each physical origin acts to generate a U(1) gauge field, 
and then present the logic of when and why the cancellation can occur and how to avoid it, 
based on the symmetry argument that applies to these noninteracting magnon Hamiltonians. 
%

\subsection{Origin of effective U(1) gauge-fields}
\label{subsec:U(1)-origin}
To elucidate the origin of U(1) gauge fields, 
it is typically enough to consider the Hamiltonian (\ref{eq:H-AFM-chain}) with dominant \textit{ferromagnetic} 
Heisenberg term and DM interactions.
Its low-energy excitation can be described using a single species of magnon, representing the 
fluctuation of ferromagnetically ordered or nearly ferromagnetically ordered moments. 
The following subsections are devoted to how each of the mechanisms works to generate the U(1) gauge fields 
on the kinetics of that single-complnent magnon. 

\begin{figure}[t]
    \centering
    \includegraphics[width=155mm]{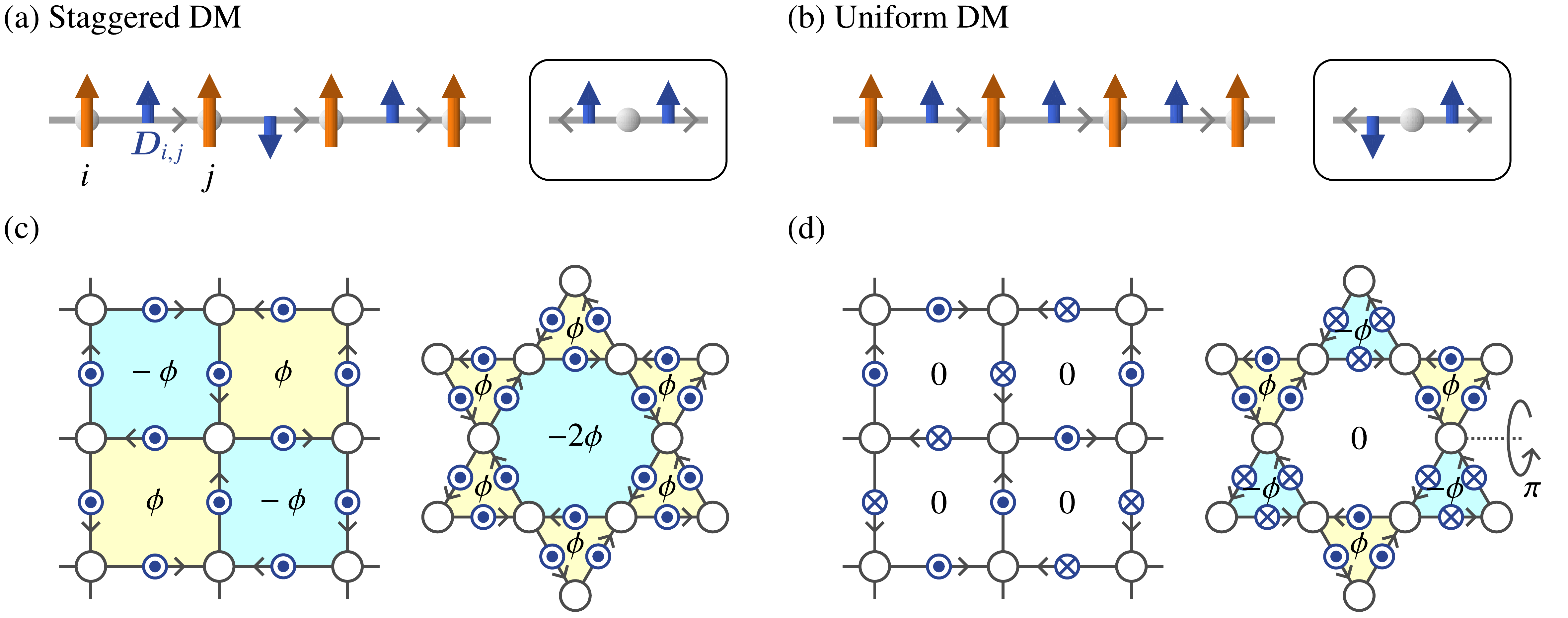}
    \caption{(a) Staggered and (b) uniform Dzyaloshinskii-Moriya (DM) interaction. The gray arrows indicate the direction of $i\to j$ in $\bm{D}_{i,j}$. In the boxes, we show the directions of the DM vectors when going from a site to its left and right neighboring sites. (c),(d) Flux pattern in ferromagnetic insulators with the (c) staggered DM and (d) uniform DM interactions. While the staggered DM interaction generates the staggered flux, the uniform DM interaction does not.}
    \label{fig:dm-alignment}
\end{figure}

\subsubsection{DM interaction}
\label{subsubsec:U(1)-DM}
\quad 
The DM interaction, $\bm{D}_{i,j}\cdot(\hat{\bm{S}}_{i}\times\hat{\bm{S}}_{j})$, 
is an antisymmetric exchange interaction that originates from SOC of magnetic ions hosting $d$ and $f$ electrons~\cite{dzyaloshinskii1958jpcs,moriya1960pr}. 
When these electrons form a multiplet 
due to the interplay of SOC with the electronic Coulomb interactions and the crystal field effect acting on the magnetic ion, 
the lowest energy multiplet separated from other levels typically forms pseudo-spin degrees of freedom consisting of both the 
spin and angular momentum. 
As a result, the lowest-order exchange interactions between the pseudo-spins on the neighboring ions include 
the Heisenberg interactions, DM interactions, and several other anisotropic exchange terms, part of which will be discussed shortly. 
\par
How these interactions appear depends on the crystal symmetry.
To have a finite DM interaction, 
the bond-centered inversion symmetry needs to be broken~\cite{dzyaloshinskii1958jpcs,moriya1960pr}. 
However, this broken symmetry can be local; even when the bond-centered inversion is broken, 
the global inversion is kept by putting the inversion center on a certain lattice site, referred to as a centrosymmetric crystal. 
Further breaking the global inversion makes the crystal noncentrosymmetric. 
Here, the U(1) gauge relevant to the insulating ferromagnets can appear 
for the former centrosymmetric case and generally not for the noncentrosymmetric ones. 
\par
The above-mentioned types of inversion-symmetries is encoded in the configuration of the DM vector, $\bm{D}_{i,j}=(D_{i,j}^{x},D_{i,j}^{y},D_{i,j}^{z})$, on the lattice bonds, which is a directional vector satisfying $\bm{D}_{i,j}=-\bm{D}_{j,i}$. 
The presence of the global inversion in centrosymmetric crystals 
means that for $1\rightarrow 2\rightarrow 3 \cdots $ along a certain direction, 
we have a staggered DM configuration, i.e., $\bm D_{1,2}=-\bm D_{2,3}=\bm D_{3,4}=\cdots$. 
For the broken global inversion in noncentrosymmetric crystals, 
$\bm D_{1,2}=\bm D_{2,3}=\cdots$, which we call uniform DM configuration. 
The schematic illustrations are shown in Fig.~\ref{fig:dm-alignment}(a) and \ref{fig:dm-alignment}(b).
Notice that the thermal Hall effect by the U(1) gauge field does not appear for the uniform DM case as we see shortly in Sec.~\ref{subsubsec:nogo} (see also Fig.~\ref{fig:dm-alignment}(c) and \ref{fig:dm-alignment}(d)).
\par
The simplest description of the U(1) gauge field is obtained from the combination of the ferromagnetic Heisenberg and the DM interactions~\cite{onose2010science}.
Here, the ground state is a ferromagnet with its spins pointing in the $+z$ direction. 
By applying Eq.~(\ref{eq:HPtr}) and neglecting magnon-magnon interactions, one finds,
\begin{align}
    -J\hat{\bm{S}}_{i}\cdot\hat{\bm{S}}_{j}+\bm{D}_{i,j}\cdot(\hat{\bm{S}}_{i}\times\hat{\bm{S}}_{j})\simeq-\sqrt{J^{2}+(D_{i,j}^{z})^{2}}S\left(\mathrm{e}^{\mathrm{i}\theta_{i,j}}\hat{b}_{i}^{\dagger}\hat{b}_{j}+\mathrm{H.c.}\right)+\cdots,
\end{align}
where $\theta_{i,j}=\mathrm{arctan}(D_{i,j}^{z}/J)$ and we dropped the constant and on-site terms. 
The DM interaction introduces a finite Peierls phase $\theta_{i,j}$ to the hopping of magnons, 
namely, acts as a vector potential or a U(1) gauge field~\cite{fujimoto2009prl,katsura2010prl,onose2010science}. 
The important remark here is that such a phase appears when and only when the DM vector has a finite component parallel to 
the spin orientation. 
Accordingly, the DM vector perpendicular to the ferromagnetic moment does not generate an effective U(1) gauge field (see Fig.~\ref{fig:u1-gauge}). 
This means that the magnon Hamiltonian restores part of the broken symmetries of the crystal or the spin Hamiltonian. 

\subsubsection{Scalar spin chirality}
\label{subsubsec:ssc}
\quad
The scalar spin chirality (SSC), $(J_{\chi}/S)\hat{\bm{S}}_{i}\cdot(\hat{\bm{S}}_{j}\times\hat{\bm{S}}_{k})$, 
is associated with the handedness of the three neighboring spins forming a triangle, 
with $i \rightarrow j \rightarrow k$ given counterclockwise and $J_{\chi}/S$ defined as a coupling constant. 
It naturally appears in systems with broken TRS~\cite{sen1995prb,wen1989prb,ye1999prl}, 
and provides another route to have an effective U(1) gauge field~\cite{katsura2010prl}. 
\par
Again we assume that the spins form a collinear ferromagnetic order. 
Applying Eq.~(\ref{eq:HPtr}) reduces the Heisenberg and the SSC terms to a magnon hopping term as
\begin{align}
    &-J(\hat{\bm{S}}_{i}\cdot\hat{\bm{S}}_{j}+\hat{\bm{S}}_{j}\cdot\hat{\bm{S}}_{k}+\hat{\bm{S}}_{k}\cdot\hat{\bm{S}}_{i})+\frac{J_{\chi}}{S}\hat{\bm{S}}_{i}\cdot(\hat{\bm{S}}_{j}\times\hat{\bm{S}}_{k})\nonumber\\
    &\quad\simeq\sqrt{J^{2}+J_{\chi}^{2}}S\left[\mathrm{e}^{\mathrm{i}\theta}(\hat{b}_{i}^{\dagger}\hat{b}_{j}+\hat{b}_{j}^{\dagger}\hat{b}_{k}+\hat{b}_{k}^{\dagger}\hat{b}_{i})+\mathrm{H.c.}\right]+\cdots,
\label{eq:scalarchiral}
\end{align}
where $\theta=\mathrm{arctan}(J_{\chi}/J)$. 
Notice that, both the Peierls phase $\theta$ and the strength of the magnon hopping do not depend on the magnitude 
of $\langle \hat{\bm{S}}_{i}\cdot(\hat{\bm{S}}_{j}\times\hat{\bm{S}}_{k}) \rangle$ itself. 
This is because the ferromagnetic moment is collinear, and only the magnon kinetics is affected by the SSC term. 
\par
The interactions other than $(J_{\chi}/S)\hat{\bm{S}}_{i}\cdot(\hat{\bm{S}}_{j}\times\hat{\bm{S}}_{k})$ can also generate a U(1) gauge field
~\cite{owerre2017prb,owerre2017epl,owerre2018prb,laurell2018prb}. 
Again suppose that the three spins form a noncoplanar structure as shown in Fig.~\ref{fig:u1-gauge}, 
and introduce three unit vectors representing their orientations as
\begin{align}
    \bm{m}_{i}&=\bm{e}^{y}\cos\zeta+\bm{e}^{z}\sin\zeta,\\
    \bm{m}_{j}&=\left(-\frac{\sqrt{3}}{2}\bm{e}^{x}-\frac{1}{2}\bm{e}^{y}\right)\cos\zeta+\bm{e}^{z}\sin\zeta,\\
    \bm{m}_{k}&=\left(\frac{\sqrt{3}}{2}\bm{e}^{x}-\frac{1}{2}\bm{e}^{y}\right)\cos\zeta+\bm{e}^{z}\sin\zeta,
\end{align}
on the global $xyz$-coordinate. 
Two limiting cases are the $\zeta=0$ representing a $120^{\circ}$ order in the $xy$-plane and $\zeta=\pi/2$ representing a ferromagnetic order. 
As the H-P transformation (\ref{eq:HPtr}) is performed not on the global $xyz$-axis but on the local $XYZ$-axis with $\bm{e}_{i}^{Z}=\bm{m}_{i}$,
for the $Y$-axes we take $\bm{e}_{i}^{Y}=\bm{e}^{x}$, $\bm{e}_{j}^{Y}=(-1/2)\bm{e}^{x}+(\sqrt{3}/2)\bm{e}^{y}$, 
and $\bm{e}_{k}^{Y}=(-1/2)\bm{e}^{x}-(\sqrt{3}/2)\bm{e}^{y}$, and $\bm e_i^X$ is determined accordingly. 
\par
At large $\zeta$ ($\simeq\pi/2$), the three spins form a canted ferromagnet,
where the magnon hopping terms $JS(\bm{e}_{i}^{+}\cdot\bm{e}_{j}^{-})\hat{b}_{i}^{\dagger}\hat{b}_{j}$ overwhelm the magnon pairing terms $JS(\bm{e}_{i}^{+}\cdot\bm{e}_{j}^{+})\hat{b}_{i}^{\dagger}\hat{b}_{j}^{\dagger}$ and the noncoplanar spin configuration generates a Peierls phase
\begin{align}
    \bm{e}_{i}^{+}\cdot\bm{e}_{j}^{-}&=\frac{3}{4}\cos^{2}\zeta-\frac{1}{2}+\mathrm{i}\frac{\sqrt{3}}{2}\sin\zeta\nonumber\\
    &\simeq\exp\left(\mathrm{i}\frac{2\pi}{3}+\mathrm{i}\frac{\sqrt{3}}{4}\sin\zeta\cos^{2}\zeta\right).
\end{align}
The phase $2\pi/3$ can be removed by a U(1) gauge transformation since it only yields a trivial phase $2\pi$ when accumulated around a triangular unit.
Then the Heisenberg exchange interaction is converted to a similar form as Eq.~(\ref{eq:scalarchiral}),
\begin{align}
    J(\hat{\bm{S}}_{i}\cdot\hat{\bm{S}}_{j}+\hat{\bm{S}}_{j}\cdot\hat{\bm{S}}_{k}+\hat{\bm{S}}_{k}\cdot\hat{\bm{S}}_{i})\simeq JS\left[\mathrm{e}^{\mathrm{i}\chi_{i,j,k}/6}(\hat{b}_{i}^{\dagger}\hat{b}_{j}+\hat{b}_{j}^{\dagger}\hat{b}_{k}+\hat{b}_{k}^{\dagger}\hat{b}_{i})+\mathrm{H.c.}\right]+\cdots,
\end{align}
where $\chi_{i,j,k}=\bm{m}_{i}\cdot(\bm{m}_{j}\times\bm{m}_{k})=(3\sqrt{3}/2)\sin\zeta\cos^{2}\zeta$ is the SSC, and is proportional to the solid angle. 
Therefore, the spin configuration with finite SCC also acts as a source of the emergent U(1) gauge field for magnons even without the explicit three-spin interaction. 

\subsubsection{Real-space spin texture}
\label{subsubsec:U(1)-spin-texture}
\quad
The scheme that the SSC with finite solid angle yielded the U(1) gauge field is extended to wider classes of spin textures
~\cite{hoogdalem2013prb,kong2013prl,iwasaki2014prb,oh2015prb,roldan2016njp,mook2017prb,kim2019prl,nikolic2020prb}. 
We illustrate it by using the triangular lattice with its ground-state spins forming a slowly varying spin texture (for example, see Fig.~\ref{fig:u1-gauge}), 
while notice that similar arguments can be applied to other lattice geometries.
To formulate such a picture requires a coarse-grained treatment using a field-theoretical scheme;
the smooth spin texture allows us to employ the continuum description, 
$\sum_{i}\to(1/v)\int\mathrm{d}^{2}\bm{r}$ and $\hat{\bm{S}}_{i}\to v\hat{\bm{s}}(\bm{r})$, where $v$ is the unit area and $\bm{r}$ 
is the positional vector. We consider a dominant ferromagnetic Heisenberg interaction and find that, 
\begin{align}
    -J\sum_{\braket{i,j}}\hat{\bm{S}}_{i}\cdot\hat{\bm{S}}_{j}+\cdots\simeq-3Jv\int\mathrm{d}^{2}\bm{r}\ \hat{\bm{s}}(\bm{r})\cdot\left(1+\frac{a^{2}}{4}\nabla^{2}\right)\hat{\bm{s}}(\bm{r})+\cdots,
    \label{eq:FM-Heisenberg-continuum}
\end{align}
where we dropped the additional terms that stabilize the given texture such as the DM interaction. 
One can also take the continuum limit of the H-P transformation as 
\begin{align}
    \hat{\bm{s}}(\bm{r})\simeq\sqrt{\frac{S}{v}}\left(\hat{b}(\bm{r})\bm{e}^{-}(\bm{r})+\mathrm{H.c.}\right)+\left(\frac{S}{v}-\hat{b}^{\dagger}(\bm{r})\hat{b}(\bm{r})\right)\bm{m}(\bm{r}),
    \label{eq:HPtr-continuum}
\end{align}
where $\bm{m}(\bm{r})$ is the unit vector pointing in the ordered spin direction. 
Appying Eq.~(\ref{eq:HPtr-continuum}) to Eq.~(\ref{eq:FM-Heisenberg-continuum}) yield the effective magnon Hamiltonian
\begin{align}
    -3Jv\int\mathrm{d}^{2}\bm{r}\ \hat{\bm{s}}(\bm{r})\cdot\left(1+\frac{a^{2}}{4}\nabla^{2}\right)\hat{\bm{s}}(\bm{r})\simeq-\frac{3}{2}JSa^{2}\int\mathrm{d}^{2}\bm{r}\ \hat{b}^{\dagger}(\bm{r})\left(\bm{\nabla}-i\bm{A}(\bm{r})\right)^{2}\hat{b}(\bm{r})+\cdots,
    \label{eq:Heff-spin-texture}
\end{align}
where $\bm{A}(\bm{r})=(A_{x}(\bm{r}),A_{y}(\bm{r}))\; (\in \mathbb{R}^{2})$ is the vector field with
\begin{align}
    A_{\mu}(\bm{r})=i\bm{e}^{+}(\bm{r})\cdot\partial_{\mu}\bm{e}^{-}(\bm{r}). 
    \label{eq:A-from-texture}
\end{align}
The form, $(\bm{\nabla}-i\bm{A}(\bm{r}))^{2}$, on the right-hand side of Eq.~(\ref{eq:Heff-spin-texture}) 
indicates that the vector field $\bm{A}(\bm{r})$ can be regarded as the vector potential, or the U(1) gauge field, for magnons.

\subsubsection{Kitaev and $\Gamma$ interaction}
\quad
The Kitaev and $\Gamma$ interactions are the typical anisotropic bond-dependent exchange interactions for the strong SOC materials
~\cite{kitaev2006ap,jackeli2009prl,chaloupka2010prl,rau2014prl}; 
$K\sum_{\gamma}\sum_{\braket{i,j}_{\gamma}}\hat{S}_{i}^{\gamma}\hat{S}_{j}^{\gamma}$ and $\Gamma\sum_{\alpha,\beta,\gamma}\sum_{\braket{i,j}_{\gamma}}(\hat{S}_{i}^{\alpha}\hat{S}_{j}^{\beta}+\hat{S}_{i}^{\beta}\hat{S}_{j}^{\alpha})$, where $\braket{i,j}_{\gamma}$ is the $\gamma$-bond ($\gamma=x,y,z$) on the honeycomb lattice as indicated in Fig.~\ref{fig:u1-gauge} and $(\alpha,\beta,\gamma)$ denotes a cyclic permutation of $(x,y,z)$. 
For simplicity, we assume $K$ and $\Gamma$ to be common to all bonds. 
\par
The thermal Hall effect of magnons in the Kitaev-Heisenberg-$\Gamma$ model is studied for the fully polarized ferromagnetic phase at high fields~\cite{mcclarty2018prb}. 
Using the perturbation expansion with respect to $1/h$, they find that the Kitaev and $\Gamma$ interactions are reduced to the on-site correction and the DM interaction between second-nearest-neighbor spin pairs, $\braket{\braket{i,j}}$, as 
\begin{align}
    (K+\Gamma)^{2}S^{2}\sum_{\braket{\braket{i,j}}}\frac{\bm{h}}{h}\cdot(\hat{\bm{S}}_{i}\times\hat{\bm{S}}_{j}),
    \label{eq:DM-from-kitaev}
\end{align}
Since the corresponding DM vector $\propto\bm{h}$ is parallel to the ground-state spins, Eq.~(\ref{eq:DM-from-kitaev}), 
the U(1) gauge field is generated as we discussed in Sec.~\ref{subsubsec:U(1)-DM}.

\subsubsection{Aharonov-Casher effect}
\quad
Magnons are by definition particles carrying a magnetic dipole moment, and thus can couple to an external electric field $\bm{E}(\bm{r})$ via the AC effect. In collinear magnetic insulators, the AC effect appears in the form of a Peierls phase~\cite{nakata2017prb1,nakata2017prb2},
\begin{align}
    \exp\left(\frac{\mathrm{i}}{\hbar}\int_{\bm{r}_{i}}^{\bm{r}_{j}}\bm{A}_{\mathrm{AC}}(\bm{r})\cdot\mathrm{d}\bm{r}\right)\hat{b}_{i}^{\dagger}\hat{b}_{j},
\end{align}
with its U(1) gauge field generated by the external electric field
\begin{align}
    \bm{A}_{\mathrm{AC}}(\bm{r})=\frac{1}{c^{2}}\bm{\mu}\times\bm{E}(\bm{r}),
\end{align}
where $c$ is the speed of light and $\bm{\mu}$ is the magnetic dipole moment of the magnon. 
In particular, a spatially inhomogeneous electric field can realize a position-dependent Peierls phase analogous to those of the Hofstadter-type models, where the electrons are subjected to the uniform external magnetic field and exhibit the conventional Hall effect. In this sense, the AC effect enables an artificial realization of the magnon thermal Hall effect on arbitrary lattices.

\begin{figure}[t]
    \centering
    \includegraphics[width=155mm]{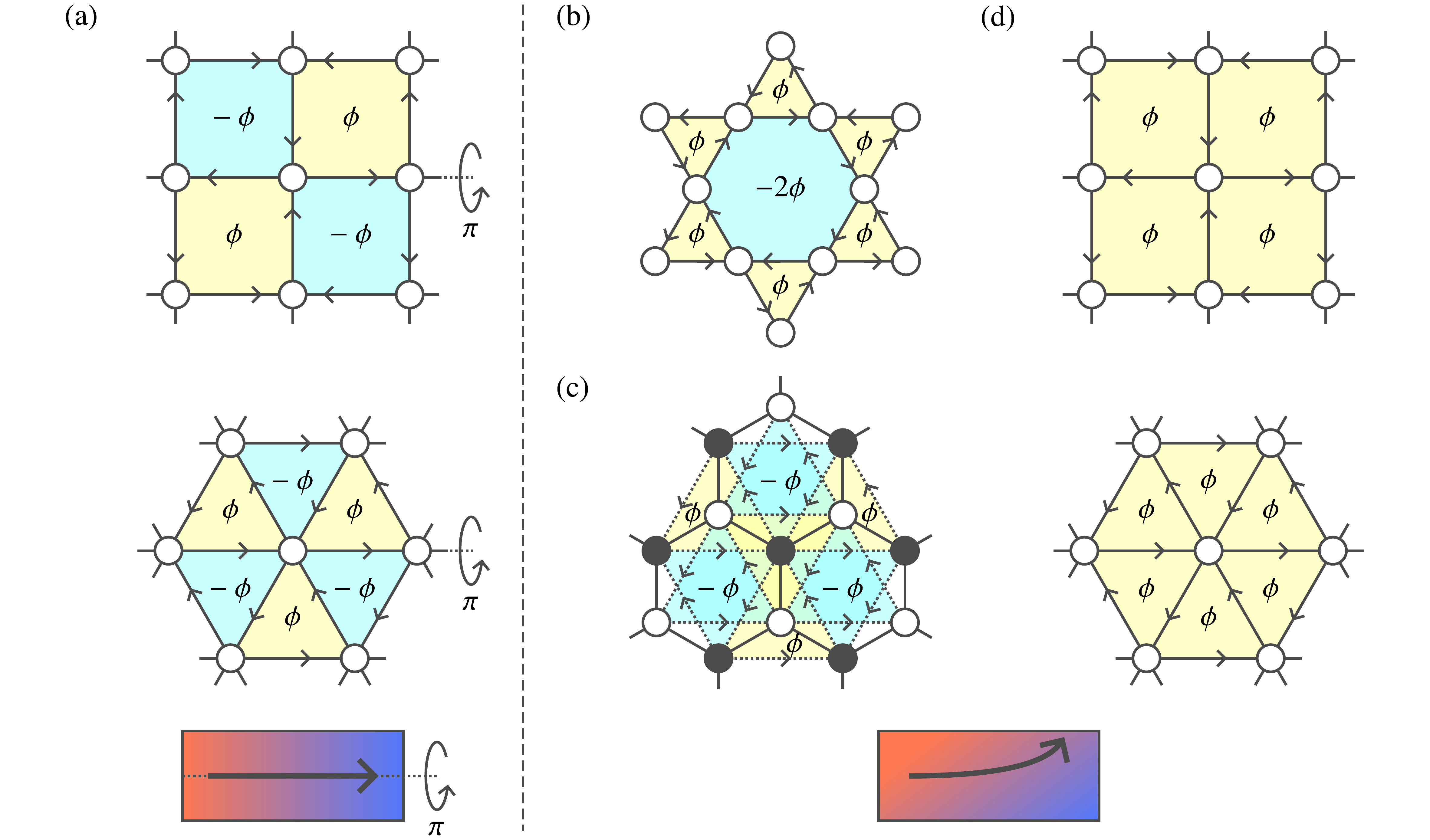}
    \caption{Schematic illustration of lattice-geometry constraints with the U(1) gauge flux. (a) Square and triangular lattices (edge-shared geometry) with the staggered flux. The flux pattern is invariant under a $\pi$ rotation about any bonds, leading to $\kappa_{xy}=0$ (bottom). (b) Kagome lattice (corner-shared geometry) with the staggered flux, (c) honeycomb lattice with the staggered flux, and (d) edge-shared lattices (square and triangular lattices) with the uniform flux. Their flux patterns no longer allow the $\pi$-rotation operation, leading to $\kappa_{xy}\neq0$ (bottom).}
    \label{fig:no-go}
\end{figure}

\subsection{Constraints from lattice geometry}
The emergence of an effective U(1) gauge field for magnons does not, by itself, guarantee the thermal Hall effect. 
As mentioned, there are three layers of discussion about the symmetry argument. 
First, when applying the LSW or equivalently the H-P transformation, we assume some sort of a magnetic ordering in the 
ground state, which, in most cases, breaks part of the symmetry of the original spin Hamiltonian. 
Second, the lowest order approximations at the noninteracting level of magnons given so far, do not necessarily 
impose the same symmetry on the effective magnon Hamiltonian as that of the ground state, 
where in some cases, the symmetry is partially restored. 
Finally, the system's symmetry after restored still includes the information of the underlying lattice geometry, 
and one needs to examine whether there exists a nonvanishing thermal Hall conductivity under the constraint from that symmetry. 
We first discuss the gauge redundancy in the magnon description to emphasize the role of gauge-invariant fluxes, 
and then classify the lattice geometries relevant to the magnon Hamiltonian. 

\subsubsection{Gauge redundancy and flux}
\label{subsubsec:gauge-invariance}
\quad
In magnon systems, the U(1) gauge redundancy naturally arises from the freedom in choosing the $XYZ$ local coordinate representing spin space. 
In our H-P transformation, the local $Z$-axis is taken parallel to the magnetic moment $\bm{m}_{i}$, 
while there remain $X,Y$ degrees of freedom that can be determined arbitrarily. 
To see this redundancy, we introduce another orthonormal basis $\{\bm{e}_{i}^{\prime X},\bm{e}_{i}^{\prime Y},\bm{e}_{i}^{Z}\}$ 
rotated from the $XY$ basis by angle $-\varphi_{i}$ about the $Z$-axis as 
\begin{align}
    \bm{e}_{i}^{\prime X}&=\bm{e}_{i}^{X}\cos\varphi_{i}-\bm{e}^{Y}\sin\varphi_{i},\\
    \bm{e}_{i}^{\prime Y}&=\bm{e}_{i}^{X}\sin\varphi_{i}+\bm{e}^{Y}\cos\varphi_{i}.
\end{align}
Introducing the corresponding magnon operator, $\hat{b}_{i}^{\prime}$, the H-P transformation is rewritten as
\begin{align}
    \hat{\bm{S}}_{i}&\simeq\sqrt{S}(\hat{b}_{i}'\bm{e}_{i}'^{-}+\mathrm{H.c.})+(S-\hat{b}_{i}'^{\dagger}\hat{b}_{i}')\bm{e}_{i}^{Z}\nonumber\\
    &=\sqrt{S}(\mathrm{e}^{-\mathrm{i}\varphi_{i}}\hat{b}_{i}'\bm{e}_{i}^{-}+\mathrm{H.c.})+(S-\hat{b}_{i}'^{\dagger}\hat{b}_{i}')\bm{e}_{i}^{Z}.       \label{eq:HPtr-gaugetr}
\end{align}
By comparing Eqs.~(\ref{eq:HPtr}) and (\ref{eq:HPtr-gaugetr}), we find
\begin{align}
    \hat{b}_{i}^{\prime}=\mathrm{e}^{\mathrm{i}\varphi_{i}}\hat{b}_{i}, 
\end{align}
which is nothing but a local U(1) gauge transformation of $\hat{b}_{i}$'s. 
The Peierls phase can formally arise even purely from the choice of the local orthonormal basis, 
meaning that there is a gauge redundancy. 
The local U(1) gauge transformation, $\hat{b}_{i}\to\mathrm{e}^{\mathrm{i}\varphi_{i}}\hat{b}_{i}$, modifies 
the hopping term as 
\begin{align}
    \hat{b}_{i}^{\dagger}\hat{b}_{j}\to\mathrm{e}^{-\mathrm{i}(\varphi_i-\varphi_j)}\hat{b}_{i}^{\dagger}\hat{b}_{j}.
\end{align}

Whereas, only the gauge-invariant quantities are relevant to physical observables that do not depend on the choice of local coordinates. 
Here, the gauge-invariant quantity relevant to the thermal Hall conductivity is the flux. 
For a given loop $C$, the flux is defined as the sum of the Peierls phases $\theta_{i,j}$ along the loop in the counterclockwise direction,
\begin{align}
    \phi_{C}=\sum_{\braket{i,j}\in C}\theta_{i,j}.
\end{align}
Under a local U(1) gauge transformation, $\theta_{i,j}\rightarrow\theta_{i,j}+\varphi_{i}-\varphi_{j}$, the phase shifts on individual bonds cancel out along the closed loop, leaving $\phi_{\mathrm{C}}$ invariant. 
Therefore, the flux pattern defined on plaquettes is physically meaningful, while the phase defined on the bonds is not. 
In the continuum limit, the rotation of the local spin coordinate by $-\varphi(\bm{r})$ generates the local U(1) gauge transformation 
of the vector field in Eq.~(\ref{eq:A-from-texture}) as
\begin{align}
    \bm{A}(\bm{r})\to\bm{A}(\bm{r})+\bm{\nabla}\varphi(\bm{r}).
\end{align}
The gauge-invariant quantity is an effective magnetic field defined as
\begin{align}
    B(\bm{r})=\partial_{x}A_{y}(\bm{r})-\partial_{y}A_{x}(\bm{r}).
    \label{eq:B-from-A}
\end{align}

\subsubsection{No-go rule for U(1) flux arrangement}
\label{subsubsec:nogo}
\quad
Let us confine ourselves to the ferromagnets or noncollinear ferromagnets which carry a finite Peierls phase 
that generates a finite flux around the closed loop. 
This condition is satisfied typically for the staggered DM interactions or SSC 
discussed in Sec.~\ref{subsubsec:U(1)-DM} and Sec.~\ref{subsubsec:ssc}. 
For such cases, the no-go rule for the thermal Hall effect of magnons about the edge-shared lattices is known~\cite{katsura2010prl,ideue2012prb}.
\par
We now illustrate this no-go rule using a square-lattice ferromagnet with nearest-neighbor Heisenberg and DM interactions, 
where all DM vectors are parallel to the ferromagnetic moment ($Z$-axis). 
As emphasized in Sec.~\ref{subsubsec:gauge-invariance}, the thermal Hall effect of magnons is governed 
by the gauge-invariant flux pattern, which is given for each square unit 
as $\pm\phi=\pm4\mathrm{arctan}(D/J)$ with $D=|\bm{D}_{i,j}|$, where we take the anticlockwise path as positive.
Importantly, there arises a staggered flux pattern for the centrosymmetric DM interaction combined with the ferromagnet, 
as illustrated in Fig.~\ref{fig:no-go}(a). Here, notice that the SSC can also generate the same flux pattern. 
Since the neighboring plaquettes have the same shape but opposite flux, the flux pattern is invariant under the $\pi$ rotation 
about one of the bands, which simultaneously exchanges the neighboring plaquettes and reverses the sign of the flux. 
Since the same $\pi$-rotation operation requires the conversion of the sign of the thermal Hall conductivity, 
$\kappa_{xy}\rightarrow -\kappa_{xy}$, these two are compatible only when $\kappa_{xy}=0$. 
The same argument applies to other edge-shared lattice geometries like the triangular lattice, 
as in their flux arrangement, the plus and minus fluxes on the neighboring triangles are related by the $\pi$ rotation operation. 
\par
Clearly, the no-go rule does not apply to the corner-shared lattices such as kagome and pyrochlore lattices. 
The kagome-lattice ferromagnet with the DM interaction illustrated in Fig.~\ref{fig:no-go}(b) 
has the opposite fluxes on the triangular and hexagonal units. 
Clearly, there is no symmetry operation that relates the two fluxes with opposite sign. 
These lattices were indeed the main platforms for the earlier studies on the magnon thermal Hall effect~\cite{katsura2010prl,onose2010science,ideue2012prb}.

\subsubsection{Honeycomb lattice}
\quad
The honeycomb lattice is edge shared but is a particular exception of the no-go rule for edge-shared lattices. 
In many magnetic insulators on the honeycomb lattice, the effective U(1) gauge field appears 
in the next-nearest-neighbor hopping processes, and the fluxes of plus and minus signs develop 
in units of triangles based on sublattices. 
Then one might naively expect that the no-go rule prohibits the thermal Hall effect of magnons. 
However, an important difference is the geometry of the lattice that does not allow the plus and minus fluxes to 
exchange by a symmetry operation; 
the $\pi$ rotation about one of the honeycomb lattice bonds is prohibited. 
Taking the $\pi$-rotation axis perpendicular to the bond exchanges the upper and lower triangles in the same hexagonal units 
(see Fig.~\ref{fig:no-go}(c)), which however carry the flux of the same sign. 
Accordingly, the lattice hosts nonvanishing thermal Hall conductivity. 

\subsubsection{Uniform flux pattern}
\quad
We now turn to the system where the fluxes of the same sign distribute widely in space (see Fig.~\ref{fig:no-go}(d)). 
Representative origin of a uniform flux is topologically-nontrivial spin texture realized in ferromagnetic skyrmion crystals (FM-SkX)
~\cite{hoogdalem2013prb,kong2013prl,iwasaki2014prb,oh2015prb,roldan2016njp,mook2017prb,kim2019prl,nikolic2020prb}. 
A slowly varying spin texture $\bm{m}(\bm{r})$ generates the effective U(1) gauge field 
and associated magnetic field $B(\bm{r})$ defined in Eq.~(\ref{eq:B-from-A}). 
From the Mermin-Ho relation~\cite{mermin1976prl}, $B(\bm{r})$ can be expressed in terms of $\bm{m}(\bm{r})$ as
\begin{align}
    B(\bm{r})=\bm{m}(\bm{r})\cdot\left(\frac{\partial\bm{m}(\bm{r})}{\partial x}\times\frac{\partial\bm{m}(\bm{r})}{\partial y}\right),
\end{align}
which is proportional to the skyrmion density. 
In the FM-SkX, the spatial average of the skyrmion density becomes finite, resulting in the uniform effective magnetic field $\bar{B}=(1/V)\int\mathrm{d}^{2}\bm{r}\ B(\bm{r})$, where $V$ is the system's area. This situation is analogous to that of charged particles subjected to a uniform magnetic field, where the Hall effect is driven by the Lorentz force. 
The uniform flux distribution thus leads to the magnon thermal Hall effect. 
There are studies proposing the engineering of similar flux patterns by the AC effect 
using spatially modulating electric fields~\cite{nakata2017prb1,nakata2017prb2}.

\section{Effective non-Abelian gauge-field picture}
\label{sec:nonabelian}
We now present another series of systems, a series of antiferromagnets, hosting the thermal Hall effect 
that can be described by a picture based on the non-Abelian gauge flux. 
Here, we mean by antiferromagnets the state that consists of two or more magnetic sublattices. 
Referring to the standard LSW formulation, each of the sublattices generates different species of magnons, 
and the fluctuations of moments due to exchange interactions between different sublattices 
generate both the exchange and the pair-creation or annihilation processes of these magnons. 
As a consequence, the magnon hopping processes are described by matrices 
with off-diagonal elements that mix different species of magnons, 
and the effective gauge fields acting on magnons are promoted from U(1) to non-Abelian ones. 
In this section, we discuss two representative examples that show the non-Abelian gauge fields. 
Then we show that the noncommutative structure of the gauge fields circumvents the no-go rules.

\begin{figure}[t]
    \centering
    \includegraphics[width=155mm]{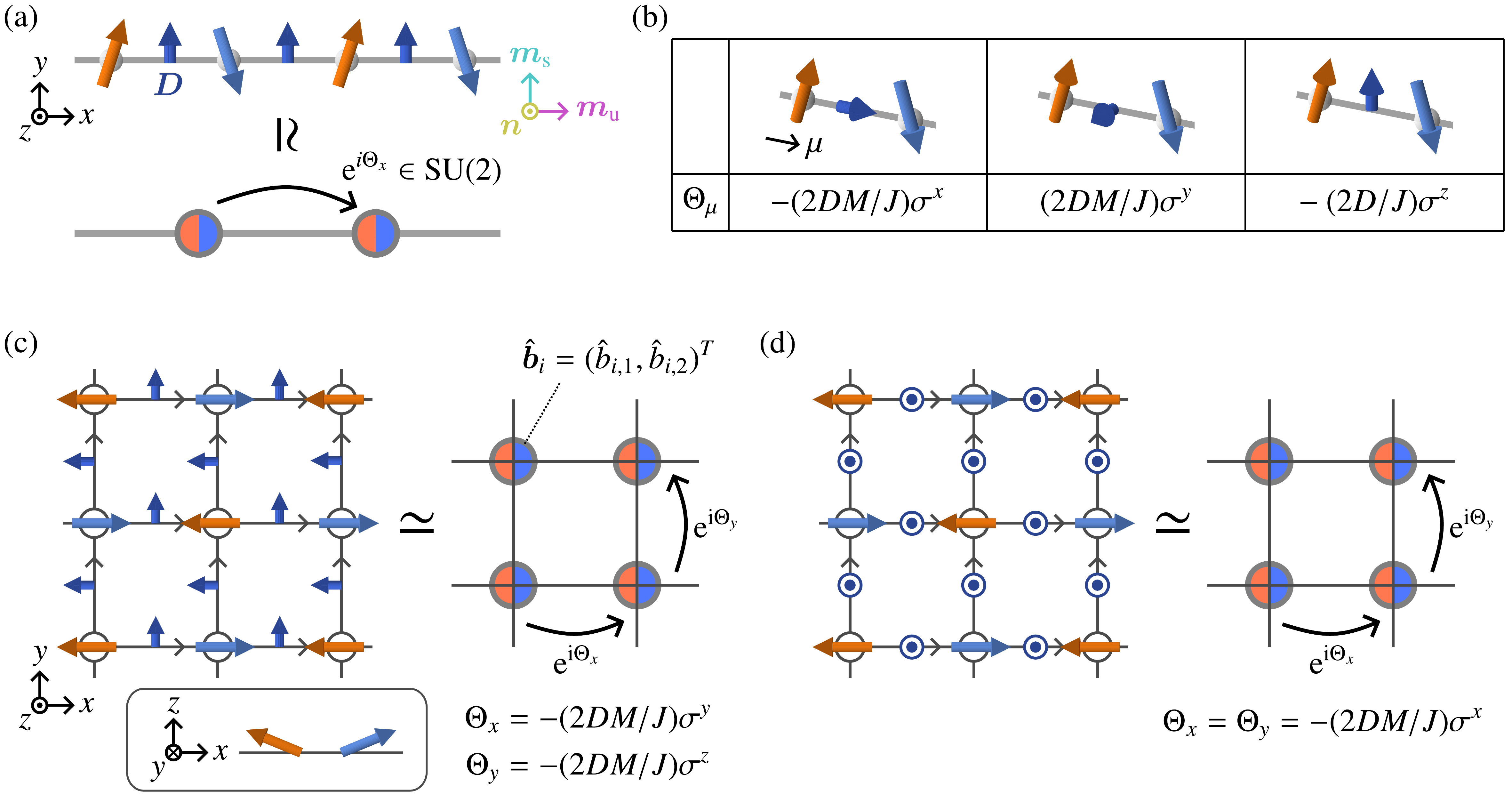}
    \caption{Effective SU(2) gauge field for magnons induced by the Dzyaloshinskii-Moriya (DM) interaction. (a) Noncollinear antiferromagnetic order with the uniform DM interaction, which is reduced to the effective lattice model with an SU(2) gauge field $\Theta_{x}$. (b) Correspondence between the noncollinear antiferromagnetic order and the associated SU(2) gauge field. (c),(d) Noncollinear antiferromagnets with the uniform DM interaction on the square lattice, which can be mapped onto the effective two-dimensional model with the SU(2) gauge field.}
    \label{fig:su2-gauge}
\end{figure}

\subsection{Two-sublattice antifeerromagnets}
\subsubsection{Model and effective field theory}
\quad
To illustrate how an effective non-Abelian gauge field is generated,
we first consider a noncollinear antiferromagnet with the DM interaction in one dimension (1D),
and the DM vectors align uniformly as shown in Fig.~\ref{fig:su2-gauge}(a).
We apply a Hamiltonian in Eq.~(\ref{eq:H-AFM-chain}), implying that the other interactions
that stabilize the noncollinear antiferromagnetic order are implicitly included,
e.g., the single-ion anisotropy, while we abbreviate them for simplicity.
\par
The fluctuation of moments on two antiferromagnetic sublattices is described using two species of magnons that behave like up and down pseudo-spin degrees of freedom.
This allows us to develop the theory for Rashba-Dresselhaus types of magnons by making an analogy 
with the electronic Rashba-Dresselhaus effect whose origin is the SOC. 
Indeed, the uniform DM interactions (see Sec.~\ref{fig:dm-alignment}(b)), whose DM vectors align uniformly along the bonds 
realized in noncentrosymetric materials, are found to be the source of the pseudo-SOC of magnons 
in the square lattice antiferromagnet~\cite{kawano2019comphys,kawano2019prb1,kawano2019prb2}. 
\par
In electronic systems, the SOC is naturally formulated in terms of an SU(2) gauge field~\cite{frohlich1993rmp}, 
and thus one can expect a similar treatment for magnons.
However, within the lattice model description, it is difficult to naturally formulate the SU(2) gauge field 
acting on two-sublattice magnons, as the pseudo-spin indices are stuck to different sublattice coordinates, 
in contrast to the electronic spins that are the internal degrees of freedom. 
This issue is resolved by introducing a coarse-grained approach we developed recently~\cite{kawano2025prb,kawano2025axv},
focusing on the long-wavelength, low-energy magnon excitations, 
for which the spatial discreteness of two magnetic sublattices becomes irrelevant.
\par
We take the continuum limit as discussed in Sec.~\ref{subsubsec:U(1)-spin-texture}, 
but now explicitly retain the sublattice degrees of freedom, 
$\sum_{i}\to(1/v')\int\mathrm{d}x\sum_{\ell}$ and $\hat{\bm{S}}_{i}\to v'\hat{\bm{s}}_{\ell}(x)$, 
where $v'$ is the area of the magnetic unit cell and $\ell=1,2$ denotes the sublattice degrees of freedom. 
The Hamiltonian (\ref{eq:H-AFM-chain}) becomes
\begin{align}
    \hat{\mathcal{H}}\simeq \int\mathrm{d}x\ \left[2Jv\hat{\bm{s}}_{1}(x)\cdot\left(1+\frac{a^{2}}{2}\partial_{x}^{2}\right)\hat{\bm{s}}_{2}(x)+2v\bm{D}\cdot\left(\hat{\bm{s}}_{1}(x)\times a\partial_{x}\hat{\bm{s}}_{2}(x)\right)\right].
\label{eq:H-AFM-chain2}
\end{align}
To specify the local orthonormal basis, we introduce the unit vectors $\bm{m}_{\mathrm{u}}$ and $\bm{m}_{\mathrm{s}}$ 
representing the uniform and staggered magnetization (see Fig.~\ref{fig:su2-gauge}(a)). 
The two unit vectors representing the noncollinear antiferromagnetic order are, 
\begin{align}
    \bm{m}_{1}&=M\bm{m}_{\mathrm{u}}+\sqrt{1-M^{2}}\bm{m}_{\mathrm{s}},\\
    \bm{m}_{2}&=M\bm{m}_{\mathrm{u}}-\sqrt{1-M^{2}}\bm{m}_{\mathrm{s}},
\end{align}
with the uniform magnetization $2M$, which is assumed to be small but finite. 
The corresponding unit vectors are, $\bm{e}_{1}^{x}=\bm{e}_{2}^{x}=\bm{n}=\bm{m}_{\mathrm{u}}\times\bm{m}_{\mathrm{s}}$ 
and $\bm{e}_{\ell}^{y}=\bm{m}_{\ell}\times\bm{n}$ ($\ell=1,2$). 
The LSW scheme in the long-wavelength regime is given by applying 
H-P transformation (\ref{eq:HPtr-continuum}) for each $\ell=1,2$ to 
Eq.~(\ref{eq:H-AFM-chain2}); 
\begin{align}
    \hat{\mathcal{H}}\simeq\frac{1}{2}\int\mathrm{d}x\ \hat{\Psi}^{\dagger}(x)H(x)\hat{\Psi}(x),
    \label{eq:Heff-magnon-continuum}
\end{align}
where $\hat{\Psi}(x)=(\hat{b}_{1}(x),\hat{b}_{2}(x),\hat{b}_{1}^{\dagger}(x),\hat{b}_{2}^{\dagger}(x))$ and $H(x)$ is the $4\times4$ matrix defined as
\begin{align}
    H(x)&=2JS(\tau^{0}\otimes\sigma^{0}+\tau^{x}\otimes\sigma^{x})+JSa^{2}\partial_{x}^{2}\tau^{x}\otimes\sigma^{x}\nonumber\\
    &\quad-i2Sa\partial_{x}\bm{D}\cdot\left[M\bm{m}_{\mathrm{u}}\tau^{z}\otimes\sigma^{x}-M\bm{n}(\tau^{x}\otimes\sigma^{y}-\tau^{0}\otimes\sigma^{y})+\bm{m}_{\mathrm{s}}\tau^{y}\otimes\sigma^{y}\right].
\end{align}
Here, we introduce unit and Pauli matrices $\tau^{\alpha}$ and $\sigma^{\alpha}$ ($\alpha=0,x,y,z$) representing the particle-hole and pseudo-spin degrees of freedom.
\par
To extract the underlying gauge structure, it is convenient to derive the Heisenberg equation of motion for magnons, $\hat{\Psi}(x,t)$, 
at time $t$ as~\cite{kawano2025prb}
\begin{align}
    -\hbar^{2}\frac{\partial}{\partial t^{2}}\hat{\Psi}(x,t)=(\tau^{z}\otimes\sigma^{0})H(x)(\tau^{z}\otimes\sigma^{0})H(x)\hat{\Psi}(x,t). 
\end{align}
In the long-wavelength limit and $|D/J|\ll1$, the $4\times4$ matrix in the right-hand side is approximated as
\begin{align}
    &(\tau^{z}\otimes\sigma^{0})H(x)(\tau^{z}\otimes\sigma^{0})H(x)\nonumber\\
    &\quad\simeq-4JS\left[JSa^{2}\partial_{x}^{2}\tau^{0}\otimes\sigma^{0}+2i\bm{D}\cdot\left(M\bm{m}_{\mathrm{u}}\tau^{z}\otimes\sigma^{x}+M\bm{n}\tau^{0}\otimes\sigma^{y}+\bm{m}_{\mathrm{s}}\tau^{z}\otimes\sigma^{z}\right)\right],
\end{align}
and finally, the equation of motion for $\hat{\bm{b}}(x,t)=(\hat{b}_{1}(x,t),\hat{b}_{2}(x,t))$ is given as
\begin{align}
    \hbar^{2}\frac{\partial}{\partial t^{2}}\hat{\bm{b}}(x,t)=(2JSa)^{2}\left(\partial_{x}\sigma^{0}-iT_{x}\right)^{2}\hat{\bm{b}}(x,t).
\end{align}
with the $2\times2$ matrix
\begin{align}
    T_{x}=-\frac{\bm{D}}{Ja}\cdot\left(M\bm{m}_{\mathrm{u}}\sigma^{x}+M\bm{n}\sigma^{y}+\bm{m}_{\mathrm{s}}\sigma^{z}\right).
    \label{eq:Tx-cotinuum}
\end{align}
Since $T_{x}$ is Hermitian and traceless, it belongs to the SU(2) Lie algebra. 
Therefore, $T_{x}$ can be interpreted as an effective SU(2) gauge field acting on the magnon pseudo-spin degrees of freedom. 
The $\sigma^{x}$ and $\sigma^{y}$ components arise from the noncollinear spin configuration ($M\neq0$), 
while the $\sigma^{z}$ component originates from the coupling between the DM interaction and the magnetic moments parallel to the DM vector.
\par
We briefly comment on the staggered DM case.
By applying the same field-theoretical arguments, one can readily show that the DM interaction does not appear in the kinetic term of the long-wavelength limit and $|D/J|\ll1$. Even for large $|D/J|$, the DM interaction simply modifies $\partial_{x}^{2}$ term and does not provide the gauge fields.

\subsubsection{1D chain as building blocks}
\quad
Let us go back to a lattice description and show how to design the system with thermal Hall effects by using the 1D chain as a building block. 
We prepare a general antiferromagnetically ordered 1D chain with a finite magnon gap, 
which is a natural assumption in the presence of a magnetic anisotropy or an external magnetic field. 
Then, the effective Hamiltonian for the low-energy magnon dynamics is given by 
\begin{align}
    \hat{\mathcal{H}}_{\mathrm{eff}}=-t_{\mathrm{eff}}\sum_{i}\left(\hat{\bm{b}}_{i}^{\dagger}\mathrm{e}^{-i\Theta_{x}}\hat{\bm{b}}_{i+1}+\mathrm{H.c.}\right)-\mu_{\mathrm{eff}}\sum_{i,\ell}\hat{b}_{i,\ell}^{\dagger}\hat{b}_{i,\ell},
    \label{eq:Heff-1Dchain}
\end{align}
where $\hat{\bm{b}}_{i}=(\hat{b}_{i,1},\hat{b}_{i,2})^{T}$ denotes the two-component magnon operator on site $i$, 
$t_{\mathrm{eff}}$ and $\mu_{\mathrm{eff}}$ are chosen to reproduce the low-energy dispersion, and
\begin{align}
    \Theta_{x}=-\frac{2\bm{D}}{J}\cdot\left(M\bm{m}_{\mathrm{u}}\sigma^{x}+M\bm{n}\sigma^{y}+\bm{m}_{\mathrm{s}}\sigma^{z}\right),
    \label{eq:Theta-mu}
\end{align}
is the SU(2) gauge field defined on the links (see Fig.~\ref{fig:su2-gauge}(a) and ~\ref{fig:su2-gauge}(b)). 
Here, one may find a good correspondence with one obtained by the continuum approach in Eq.~(\ref{eq:Tx-cotinuum}).
\par
We now combine this 1D chain and form a 2D two-sublattice antiferromagnet. 
The first example is the square-lattice antiferromagnets with the Rashba-type DM interaction shown in Fig.~\ref{fig:su2-gauge}(c); 
the DM vector respects the $C_{4v}$ point group symmetry, 
and we assume that the noncollinear antiferomagnetic order is realized in the $xz$-plane having a small uniform magnetization in the $z$-direction, namely $\bm{m}_{\mathrm{u}}=\bm{e}^{z}$ and $\bm{m}_{\mathrm{s}}=-\bm{e}^{x}$. 
The system can be regarded as the combination of the 1D chain along the $x$ and $y$ direction described by Eqs.~(\ref{eq:Heff-1Dchain}) and (\ref{eq:Theta-mu}). Then, without the need for detailed field-theoretical calculation, 
the SU(2) gauge fields along the $x$ and $y$ direction is obtained as
\begin{align}
    \Theta_{x}=-(2DM/J)\sigma^{y}, \quad \Theta_{y}=-(2DM/J)\sigma^{z}.
    \label{eq:SU2gauge-noncommutative}
\end{align}
The second example is the same square-lattice antiferromagnet, but now all the DM vectors point in the $z$-direction as illustrated in Fig.~\ref{fig:su2-gauge}(d).
In this case, the SU(2) gauge fields along the $x$ and $y$ directions are identical and is obtained 
in a similar manner as the combination of the 1D chain with uniform DM as
\begin{align}
    \Theta_{x}=\Theta_{y}=-(2DM/J)\sigma^{x}.
    \label{eq:SU2gauge-commutative}
\end{align}
\par
The similar treatment applies to the honeycomb lattice by regarding the zigzag chain 
consisting of two different bond directions~\cite{kawano2019comphys}. 
Note that the usage of $\Theta_{x/y}$ is not to obtain a quantitatively accurate evaluation of the thermal Hall conductivity, 
but to judge whether we have room to obtain a nonvanishing thermal Hall effect and how the model parameters 
are to be tuned to enhance the conductivity. 
\par
In reality, there are indeed more complicated model Hamiltonians with anisotropic spin exchange interactions 
or the staggered external fields designed on top of these lattices, and 
to know how each of the model parameters functions to have topological properties in the energy bands 
is not straightforward compared to the case of U(1) gauge fields. 
For such a case, there is a framework to judge, using the Brillouin-Wigner method, that exactly and efficiently reduces
the degrees of freedom to identify the pseudo-SOC of magnons and associated momentum-dependent magnon spin textures~\cite{kawano2019prb2}.
This treatment shows that pseudo-SOC picture straightforwardly applies to Kitaev, Gamma, and Kane-Mele-type models, 
and possibly to other models still unexplored.

\begin{figure}[t]
    \centering
    \includegraphics[width=155mm]{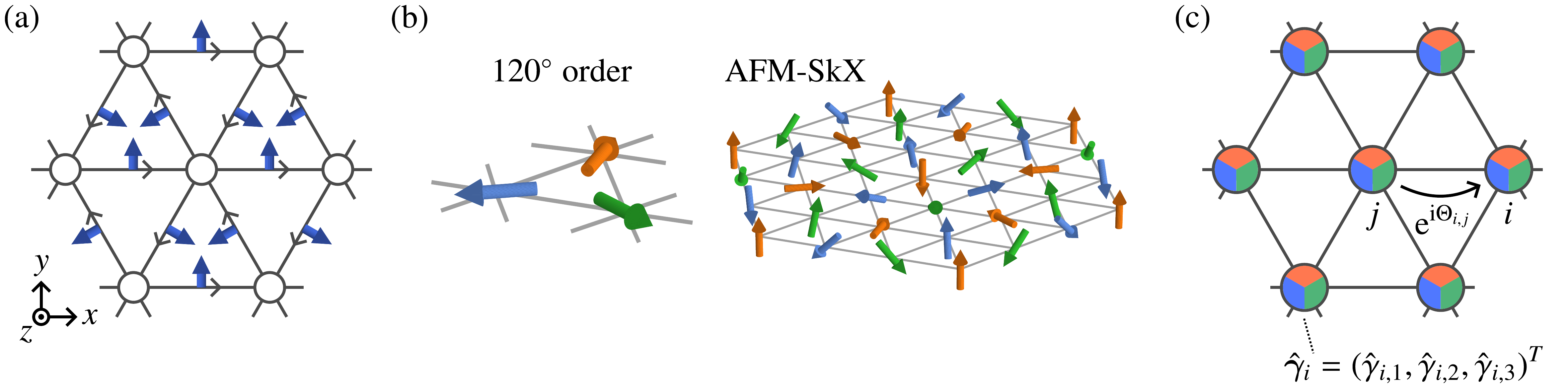}
    \caption{Effective SU(3) gauge field for magnons in three-sublattice antiferromagnets. (a) Triangular-lattice antiferromagnet with the uniform Dzyaloshinskii-Moriya (DM) interaction. The blue and gray arrows indicate the DM vector and the direction of $i\to j$ in $\bm{D}_{i,j}$. (b) Two representative three-sublattice spin configurations: coplanar $120^{\circ}$ order and antiferromagnetic skytrmion crystals. (c) Effective lattice model for three-component magnons, where the hopping is described by $3\times3$ matrix $\exp[\mathrm{i}\Theta_{i,j}]\in\mathrm{SU}(3)$.}
    \label{fig:su3-gauge}
\end{figure}

\subsection{Three-sublattice antiferromagnet}
\quad
A natural extension of the SU(2) gauge field to a higher-rank one is to increase the number of sublattices. 
As a good example relevant to experimental findings, 
we consider three-sublattice antiferromagnets, whose gauge structure, if present, should be promoted from SU(2) to SU(3). 
The Heisenberg antiferromagnet on the triangular lattice with the uniform DM interaction 
where the DM vectors lie in the $xy$ plane is shown in Fig.~\ref{fig:su3-gauge}(a), 
for which we assume that the Hamiltonian in the form of Eq.~(\ref{eq:H-AFM-chain}) with 
some extra terms to stabilize the three-sublattice magnetic ordering; 
it includes not only the $120^{\circ}$ order but also the AFM-SkX,
which is regarded as the combination of the FM-SkX on three different sublattices coupled with other sublattices (see Fig.~\ref{fig:su3-gauge}(b))~\cite{rosales2015prb,diaz2019prl,mukherjee2021scirep,mohylna2022prb}. 
The AFM-SkX is indeed realized in MnSc$_{2}$S$_{4}$ in a finite range of magnetic field, 
which consists of a stacking of triangular layers forming a diamond lattice~\cite{reil2002jac,fritsch2004prl,gao2017nphys,gao2020nature}.
We have shown that $\kappa_{xy}\neq0$ for this material in Ref.~\cite{takeda2024ncom} 
both in experiment and in theory.
\par
By expanding the effective field theory (see Ref.~\cite{kawano2025prb} for details), 
we straightforwardly develop a treatment similar to the SU(2) case.
Taking the continuum limit of the spin operators with three sublattice indices, $\hat{\bm{S}}_{i}\to v'\hat{\bm{s}}_{\ell}(\bm{r})$ ($\ell=1,2,3$) and applying the H-P transformation (\ref{eq:HPtr-continuum}) lead to the quadratic magnon Hamiltonian in terms of $\hat{b}_{\ell}(\bm{r})$ and $\hat{b}_{\ell}^{\dagger}(\bm{r})$.
However, the resulting effective Hamiltonian is complicated, making it difficult to elucidate the gauge structure. To clarify the underlying gauge structure, we introduce the new bosonic operators $\hat{\gamma}_{n}(\bm{r})$ as the linear combination of $\hat{b}_{\ell}(\bm{r})$ and $\hat{b}_{\ell}^{\dagger}(\bm{r})$
\begin{align}
    \hat{\gamma}_{1}(\bm{r})&=\frac{1}{\sqrt{3}}\sum_{\ell}\left(\cosh\chi\hat{b}_{\ell}(\bm{r})+\sinh\chi\hat{b}_{\ell}^{\dagger}(\bm{r})\right),\\
    \hat{\gamma}_{2}(\bm{r})&=i\sqrt{\frac{2}{3}}\sum_{\ell}\cos\theta_{\ell}\left(\cosh\chi\hat{b}_{\ell}(\bm{r})+\sinh\chi\hat{b}_{\ell}^{\dagger}(\bm{r})\right),\\
    \hat{\gamma}_{3}(\bm{r})&=i\sqrt{\frac{2}{3}}\sum_{\ell}\sin\theta_{\ell}\left(\cosh\chi\hat{b}_{\ell}(\bm{r})+\sinh\chi\hat{b}_{\ell}^{\dagger}(\bm{r})\right),
\end{align}
with $\theta_{\ell}=2\pi(\ell-1)/3$ and $\chi=(1/2)\mathrm{arctan}(1/3)$. Then the effective Hamiltonian becomes
\begin{align}
    \hat{\mathcal{H}}\simeq\frac{1}{2}\int\mathrm{d}^{2}\bm{r}\ \hat{\bm{\Psi}}^{\dagger}(\bm{r})H(\bm{r})\hat{\bm{\Psi}}(\bm{r}),
\end{align}
where $\hat{\bm{\Psi}}(\bm{r})=(\hat{\gamma}_{1}(\bm{r}),\hat{\gamma}_{1}^{\dagger}(\bm{r}),\hat{\gamma}_{2}(\bm{r}),\hat{\gamma}_{2}^{\dagger}(\bm{r}),\hat{\gamma}_{3}(\bm{r}),\hat{\gamma}_{3}^{\dagger}(\bm{r}))^{T}$ and $H(\bm{r})$ is the $6\times6$ matrix defined as
\begin{align}
    H(\bm{r})=\frac{3\sqrt{2}}{8}JSa^{2}\left(\tilde{I}\otimes\tau^{x}\nabla^{2}-i\bm{G}(\bm{r})\cdot\bm{\nabla}\right)+\frac{9\sqrt{2}}{4}JSI\otimes(\tau^{0}+\tau^{x}),
\end{align}
with $I=\mathrm{diag}(1,1,1)$, $\tilde{I}=\mathrm{diag}(2,1,1)$, and
\begin{align}
    \bm{G}(\bm{r})&=\bm{A}_{\mathrm{c}}(\bm{r})\left[\lambda_{2}\otimes(\tau^{0}+3\tau^{x})+\frac{1}{\sqrt{2}}(\lambda_{3}-\sqrt{3}\lambda_{8})\otimes\tau^{z}\right]\nonumber\\
    &\quad+\bm{A}_{\mathrm{s}}(\bm{r})\left[\lambda_{5}\otimes(\tau^{0}+3\tau^{x})+\sqrt{2}\lambda_{6}\otimes\tau^{z}\right]-\sqrt{3}\bm{\xi}(\bm{r})\lambda_{7}\otimes(\tau^{0}+\tau^{x}).
\end{align}
Here, $\lambda_{\alpha}$ ($\alpha=1,2,\ldots,8$) is the Gell-Mann matrix, $\bm{A}_{\mathrm{c}}(\bm{r})$ and $\bm{A}_{\mathrm{s}}(\bm{r})$ are the linear combination of the effective U(1) gauge field on each sublattice
\begin{align}
    \bm{A}_{\mathrm{c}}(\bm{r})&=\frac{1}{3}\sum_{\ell}\bm{A}_{\ell}(\bm{r})\cos\theta_{\ell},
    \label{eq:Ac}\\
    \bm{A}_{\mathrm{s}}(\bm{r})&=\frac{1}{3}\sum_{\ell}\bm{A}_{\ell}(\bm{r})\sin\theta_{\ell},
    \label{eq:As}
\end{align}
$\bm{A}_{\ell}(\bm{r})=(A_{\ell,x}(\bm{r}),A_{\ell,y}(\bm{r}))$ ($\ell=1,2,3$) is defined as
\begin{align}
    A_{\ell,x}(\bm{r})&=i\bm{e}_{\ell}^{+}(\bm{r})\cdot\partial_{x}\bm{e}^{-}(\bm{r})+(2D/Ja)m_{\ell}^{Y}(\bm{r}),
    \label{eq:A-ell-x}\\
    A_{\ell,y}(\bm{r})&=i\bm{e}_{\ell}^{+}(\bm{r})\cdot\partial_{y}\bm{e}^{-}(\bm{r})-(2D/Ja)m_{\ell}^{X}(\bm{r}),
    \label{eq:A-ell-y}
\end{align}
where $\bm{m}_{\ell}(\bm{r})$ is the unit vector in the direction of a moment on the $\ell$-th sublattice at location $\bm{r}$, and
\begin{align}
    \bm{\xi}(\bm{r})=\frac{1}{3}\sum_{(\ell,\ell')=(1,2),(2,3),(3,1)}[\bm{m}_{\ell}(\bm{r})]^{T}\bm{\nabla}\bm{m}_{\ell'}(\bm{r}).
\end{align}
As in the case of two-sublattice antiferromagnets, the equation of motion for $\hat{\bm{\gamma}}(\bm{r})=(\hat{\gamma}_{1}(\bm{r}),\hat{\gamma}_{2}(\bm{r}),\hat{\gamma}_{3}(\bm{r}))$ can be obtained as
\begin{align}
    \hbar^{2}\frac{\partial^{2}}{\partial t^{2}}\hat{\bm{\gamma}}(\bm{r},t)=\frac{27}{8}(JSa)^{2}\tilde{I}\left(\bm{\nabla}I-i\bm{T}\right)^{2}\hat{\bm{\gamma}}(\bm{r},t)+\cdots,
    \label{eq:eom-triangularAFM}
\end{align}
where
\begin{align}
    \tilde{I}\bm{T}(\bm{r})&=\bm{A}_{\mathrm{c}}(\bm{r})\left[\lambda_{2}-\frac{\sqrt{2}}{4}(\lambda_{3}-\sqrt{3}\lambda_{8})\right]+\bm{A}_{\mathrm{s}}(\bm{r})\left(\lambda_{5}-\frac{\lambda_{6}}{\sqrt{2}}\right).
    \label{eq:T(r)}
\end{align}
The Hermitian and traceless $3\times3$ matrix $T_{\mu}(\bm{r})$ ($\mu=x,y$) is intepreted as an effective SU(3) gauge field for magnons. 
Eqs.~(\ref{eq:A-ell-x}) and (\ref{eq:A-ell-y}) vidualize that 
the origin of a nonvanishing $\bm{T}(\bm{r})$ is the DM vector or a spin texture. 
The physical implications of this SU(3) gauge structure will be discussed in more detail later.
\par
We finally show the effective lattice model that reproduces the same low-energy behavior. 
As in the previous SU(2) case, we assume a finite magnon gap, and the effective Hamiltonian for 
the magnons, $\hat{\bm{\gamma}}_{i}=(\hat{\gamma}_{i,1},\hat{\gamma}_{i,2},\hat{\gamma}_{i,3})^{T}$, is written as
\begin{align}
    \hat{\mathcal{H}}_{\mathrm{eff}}=-t_{\mathrm{eff}}\sum_{\braket{i,j}}\left(\hat{\bm{\gamma}}_{i}^{\dagger}\mathrm{e}^{-i\Theta_{i,j}}\hat{\bm{\gamma}}_{j}+\mathrm{H.c.}\right)-\mu_{\mathrm{eff}}\sum_{i,n}\hat{\gamma}_{i,n}^{\dagger}\hat{\gamma}_{i,n}. 
\end{align}
The SU(3) gauge field is encoded in the $3\times3$ matrix $\Theta_{i,j}$ as (see also Fig.~\ref{fig:su3-gauge}(c))
\begin{align}
    \Theta_{i,j}=\int_{\bm{r}_{i}}^{\bm{r}_{j}}\bm{T}(\bm{r})\cdot\mathrm{d}\bm{r}.
\end{align}

\begin{figure}[t]
    \centering
    \includegraphics[width=155mm]{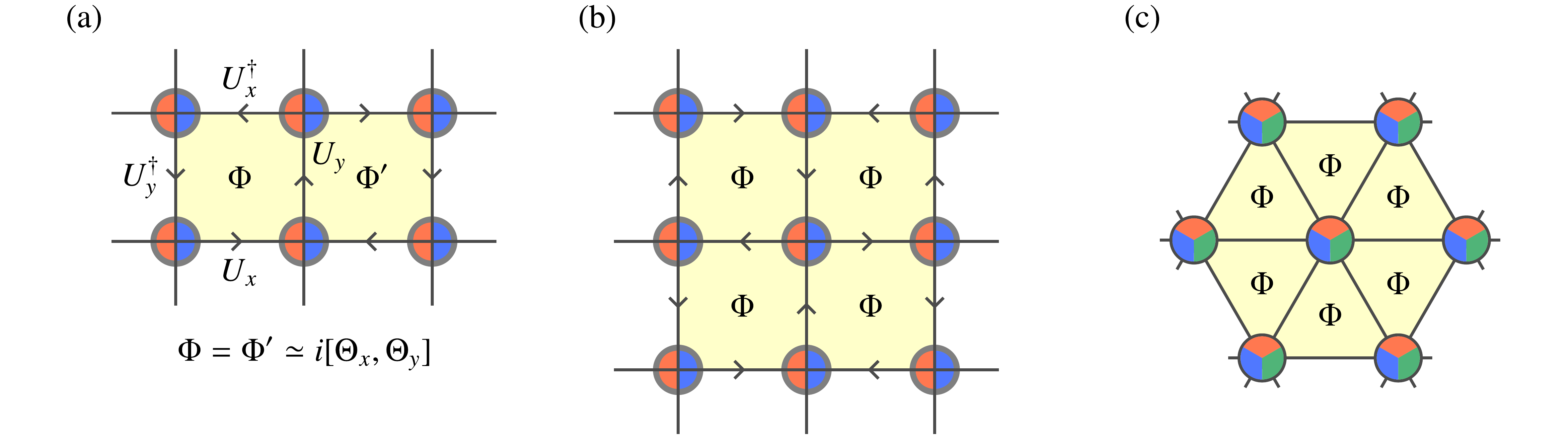}
    \caption{Circumventing the no-go rule by non-Abelian gauge fields. (a) Square lattice with the SU(2) gauge field represented by $2\times2$ matrix $U_{\mu}$ ($\mu=x,y$). The fluxes $\Phi$ and $\Phi'$ defined on an elementary plaquette in the counterclockwise and clockwise directions take the same value $\Phi=\Phi'\simeq \mathrm{i}[\Theta_x,\Theta_y]$, which originates from the noncommutativity of the gauge fields. (b) Resulting flux pattern: unlike the U(1) case, the uniform component of the flux generated by the commutator breaks the $\pi$-rotation symmetry about the bonds, leading to finite $\kappa_{xy}$. (c) Triangular lattice with the SU(3) gauge field and its flux pattern. The coplanar $120^{\circ}$ order or the antiferromagnetic skyrmion crystals generate finite $\Phi$.}
    \label{fig:avoid-no-go}
\end{figure}

\subsection{Non-abelian rule-to-go}
In the U(1) gauge field picture, the flux cancellation due to the symmetry of the magnon Hamiltonian was crucial, 
in particular, for the ferromagnets realized on edge-shared lattices hosting staggered U(1) fluxes, 
referred to as the no-go rule of the thermal Hall effect. 
However, even in these edge-shared lattice geometries, where there are several symmetry operations 
that can easily zero out the U(1) flux contribution, 
the non-Abelian ones can easily survive and circumvent the no-go rule. 
In the following, we give a hand-weaving explanation on how the symmetry constraints is lifted on the edge-shared lattices. 
\par
As an example, we consider an antiferromagnetic insulator on the square lattice. 
In the corresponding effective lattice model, the hopping term takes the form of $2\times2$ matrix, 
$U_{\mu}=\exp(-\mathrm{i}\Theta_{\mu})$ ($\mu=x,y$). 
As in the case of the U(1) gauge field, we introduce a ``flux'' on each plaquette. 
The flux $\Phi$ ($\Phi'$) around an elementary plaquette in the counterclockwise (clockwise) direction is defined as (see Fig.~\ref{fig:avoid-no-go}(a))
\begin{align}
    \mathrm{e}^{i\Phi}&=U_{x}U_{y}^{\dagger}U_{x}^{\dagger}U_{y},\\
    \mathrm{e}^{-i\Phi'}&=U_{x}^{\dagger}U_{y}^{\dagger}U_{x}U_{y},
\end{align}
and $\Phi=\Phi'\simeq\mathrm{i}[\Theta_{x},\Theta_{y}]$, which comes from the noncommutativity of the gauge field. 
Here, importantly, the flux produces the nonzero uniform component, which means that they cannot vanish by any of the symmetries~\cite{comment-su2}. 
The example shown in Fig.~\ref{fig:su2-gauge}(c) has a noncommutative SU(2) gauge field (\ref{eq:SU2gauge-noncommutative}), 
which gives a finite flux $\Phi=\Phi'=2(2DM/J)^{2}\sigma^{x}$, and thus a nonvanishing thermal Hall conductivity. 
Whereas in the system shown in Fig.~\ref{fig:su2-gauge}(d), the SU(2) gauge field (\ref{eq:SU2gauge-commutative}) along the $x$ and $y$ 
direction commutes and it does not produce a finite flux $\Phi=0$. 
One can thus judge immediately by the commutativity/noncommutativity of the gauges on whether the system shows a thermal Hall effect.
\par
We also give a complementary viewpoint from the effective field theory. 
From the SU(N) gauge field $\bm{T}(\bm{r})$, we can define the field strength, which is the extension of the magnetic field 
in Eq.~(\ref{eq:B-from-A}), as 
\begin{align}
    F_{xy}(\bm{r})=\partial_{x}T_{y}(\bm{r})-\partial_{y}T_{x}(\bm{r})-i[T_{x}(\bm{r}),T_{y}(\bm{r})].
    \label{eq:fxy}
\end{align}
In particular, the third term $-i[T_{x}(\bm{r}),T_{y}(\bm{r})]$ originates from the noncommutativity of the gauge field. 
Evaluating the strength of $F_{xy}(\bm r)$ gives a clue to know how much the thermal Hall conductivity can be enhanced 
to the realistic order that can be captured by the experiment. 
\par
In the case of triangular-lattice antiferromagnets with the $120^{\circ}$ order and the DM interaction, 
the contribution is solely from the thired term, $-i[T_{x}(\bm{r}),T_{y}(\bm{r})]$, and is calculated 
from Eqs.~(\ref{eq:T(r)})-(\ref{eq:A-ell-y}), where we find a spatially uniform value, $\bar{F}_{xy}=(1/V)\int\mathrm{d}^{2}\bm{r}F_{xy}(\bm{r})$, as~\cite{kawano2025prb}
\begin{align}
    \bar{F}_{xy}=-\frac{\sqrt{3}}{12}\left(\frac{2D}{Ja}\right)^{2}\lambda_{7}. 
    \label{eq:fxy-120-DM}
\end{align}
Therefore, the $120^{\circ}$ order with the DM interaction in the $xy$ plane exhibits a finite thermal Hall effect. 
A similar situation is demonstrated in Sec.~\ref{subsec:application}. 
\par
The AFM-SkX case also has the contribution only from the third term of Eq.~(\ref{eq:fxy}). 
Although it may seem at a glance that the first two terms that generate a U(1) gauge in each sublattice, 
their contributions cancel out due to the no-go rule as in the ferromagnets; 
the simple understanding is that each sublattice can be viewed as a system with a giant spin moment, 
and they are related to each other by symmetry operations.
The calculation of $F_{xy}$ itself is formally straightforward by substituting Eq.~(\ref{eq:T(r)}) into Eq.~(\ref{eq:fxy}). 
While, we will provide an easier evaluation for this case, obtained independently using a field-theoretical treatment~\cite{kawano2025prb}, 
\begin{align}
    \bar{F}_{xy}=-\frac{1}{8V}\int\mathrm{d}^{2}\bm{r}\ \bm{n}(\bm{r})\cdot[\partial_{x}\bm{n}(\bm{r})\times\partial_{y}\bm{n}(\bm{r})],
\end{align}
and $\bar{F}_{xy}$ is proportional to the skyrmion density of the vector field $\bm{n}(\bm{r})$, which is perpendicular to the local $120^{\circ}$ order in the AFM-SkX and also forms the topologically-nontrivial texture. Therefore, the uniform element of the field strength from the AFM-SkX spin texture is finite.

\begin{figure}[t]
    \centering
    \includegraphics[width=155mm]{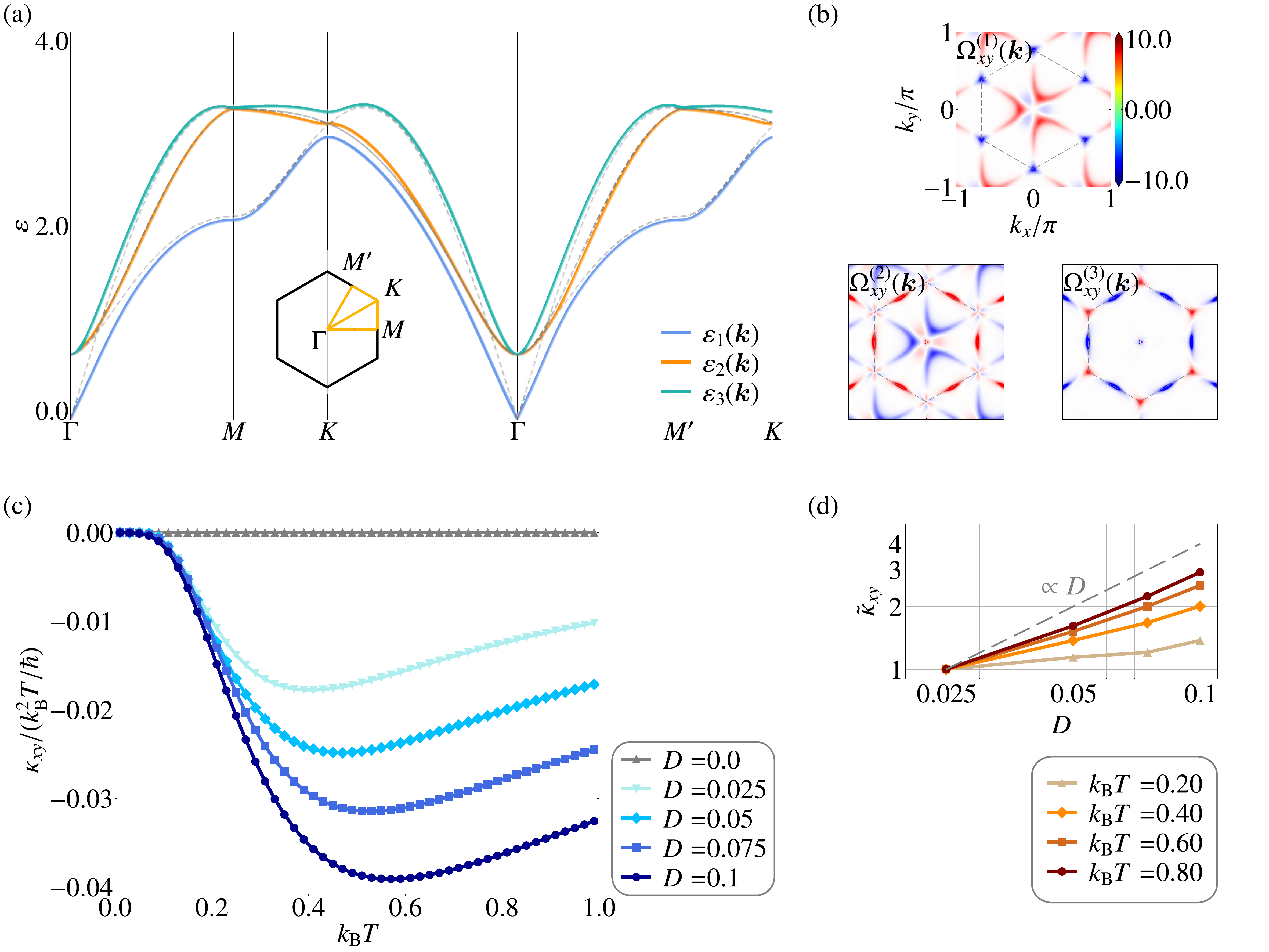}
    \caption{Magnon bands, Berry curvature, and thermal Hall conductivity. The parameters are set to $JS=1.0$ and $\Lambda=0.05$. (a) Magnon bands along the high-symmetry line in the Brillouin zone (inset) for $D=0.1$. The dashed lines denote the magnon bands with $D=0$. (b) Density plots of the Berry curvature for $D=0.1$. The dashed hexagon indicates the Brillouin zone. (c) Temperature dependence of the thermal Hall conductivity for several values of $D$. (d) $D$ dependence of the thermal Hall conductivity normalized by its value at $D=0.025$, $\tilde{\kappa}_{xy}=\kappa_{xy}(D)/\kappa_{xy}(D=0.025)$. The dashed line indicates the linear dependence on $D$ for comparison.}
    \label{fig:application}
\end{figure}

\subsection{Minimal model: triangular-lattice antiferromagnet with the uniform DM interaction}
\label{subsec:application}
\quad
We finally propose a canonical and minimal model exhibiting the thermal Hall effect due to a higher-rank gauge field, 
much simpler than the ones we have discussed in the previous subsections. 
The platform is the triangular lattice hosting a simple coplanar $120^{\circ}$ order, 
which is, in the U(1) gauge picture, subject to the no-go rule. 
Indeed, the triangular lattice systems that exhibited the thermal Hall effect 
were linited to topologically-nontrivial spin textures such as FM-SkXs and AFM-SkXs, or in some distorted lattices (see Table~\ref{table:classification}). 
\par
Our spin Hamiltonian is given as 
\begin{align}
    \hat{\mathcal{H}}=\sum_{\braket{i,j}}\left[J\hat{\bm{S}}_{i}\cdot\hat{\bm{S}}_{j}+\bm{D}_{i,j}\cdot(\hat{\bm{S}}_{i}\times\hat{\bm{S}}_{j})\right]+\Lambda\sum_{i=1}^{N}(\hat{S}_{i}^{z})^{2},
\end{align}
which consists of Heisenberg interaction $J>0$, DM interaction with $\bm{D}_{i,j}=D\bm{e}^{z}\times(\bm{r}_{j}-\bm{r}_{i})/|\bm{r}_{j}-\bm{r}_{i}|$, and single-ion magnetic anisotropy $\Lambda>0$. 
For sufficiently large $\Lambda$, the ground state exhibits the $120^{\circ}$ order. 
From the viewpoint of the SU(3) gauge-field description, the present model shows the finite field strength $\bar{F}_{xy}$ as shown in Eq.~(\ref{eq:fxy-120-DM}) and is expected to exhibit the thermal Hall effect of magnons. 
\par
Applying the H-P transformation (\ref{eq:HPtr}), we obtain a noninteracting magnon Hamiltonian. 
The Fourier transformation is given for $\ell=1,2,3$ sublattices as, 
$\hat{b}_{i}=(N/3)^{-1/2}\sum_{\bm{k}}\hat{b}_{\bm{k},\ell}\mathrm{e}^{\mathrm{i}\bm{k}\cdot\bm{r}}$, 
where $\bm{k}=(k_{x},k_{y})$ is the wave vector in the first Brillouin zone. 
By the Fourier transformation, the magnon Hamiltonian is written in the Bogoliubov-de Gennes (BdG) form as
\begin{align}
    \hat{\mathcal{H}}\simeq\frac{1}{2}\sum_{\bm{k}}\hat{\Psi}_{\bm{k}}^{\dagger}H_{\mathrm{BdG}}(\bm{k})\hat{\Psi}_{\bm{k}},
\end{align}
where $\hat{\Psi}_{\bm{k}}=(\hat{b}_{\bm{k},1},\hat{b}_{\bm{k},2},\hat{b}_{\bm{k},3},\hat{b}_{\bm{k},1}^{\dagger},\hat{b}_{\bm{k},2}^{\dagger},\hat{b}_{\bm{k},3}^{\dagger})^{T}$ and $H_{\mathrm{BdG}}(\bm{k})$ is the $6\times6$ Hermitian matrix of the form
\begin{align}
    H_{\mathrm{BdG}}(\bm{k})=
    \begin{pmatrix}
        \Xi(\bm{k}) & \Delta(\bm{k})\\
        \Delta^{\dagger}(\bm{k}) & \Xi^{*}(-\bm{k})
    \end{pmatrix}
    .
\end{align}
The normal and anomalous blocks in $H_{\mathrm{BdG}}(\bm{k})$ are given by
\begin{align}
    \Xi(\bm{k})&=
    \begin{pmatrix}
        3JS+\Lambda S & \sum_{\nu}t_{1,\nu}\mathrm{e}^{ik_{\nu}} & \sum_{\nu}t_{3,\nu}^{*}\mathrm{e}^{-ik_{\nu}}\\
        \sum_{\nu}t_{1,\nu}^{*}\mathrm{e}^{-ik_{\nu}} & 3JS+\Lambda S & \sum_{\nu}t_{2,\nu}\mathrm{e}^{ik_{\nu}}\\
        \sum_{\nu}t_{3,\nu}\mathrm{e}^{ik_{\nu}} & \sum_{\nu}t_{2,\nu}^{*}\mathrm{e}^{-ik_{\nu}} & 3JS+\Lambda S
    \end{pmatrix}
    ,\\
    \Delta(\bm{k})&=
    \begin{pmatrix}
        \Lambda S & \sum_{\nu}\Delta_{1,\nu}\mathrm{e}^{ik_{\nu}} & \sum_{\nu}\Delta_{3,\nu}\mathrm{e}^{-ik_{\nu}}\\
        \sum_{\nu}\Delta_{1,\nu}\mathrm{e}^{-ik_{\nu}} & \Lambda S & \sum_{\nu}\Delta_{2,\nu}\mathrm{e}^{ik_{\nu}}\\
        \sum_{\nu}\Delta_{3,\nu}\mathrm{e}^{ik_{\nu}} & \sum_{\nu}\Delta_{2,\nu}\mathrm{e}^{-ik_{\nu}} & \Lambda S
    \end{pmatrix}
    ,
\end{align}
where the hopping amplitudes $t_{\ell,\nu}$ and pairing terms $\Delta_{\ell,\nu}$ ($\ell,\nu=1,2,3$) are defined as
\begin{align}
    t_{\ell,\nu}&=\frac{1}{4}JS+i\frac{S}{2}\bm{D}_{\nu}\cdot\bm{d}_{\ell,\ell+1},\\
    \Delta_{\ell,\nu}&=\frac{3}{4}JS+i\frac{\sqrt{3}}{2}S\bm{D}_{\nu}\cdot(\bm{e}^{z}\times\bm{d}_{\ell,\ell+1}),
\end{align}
and $\bm{d}_{\ell,\ell+1}=(\bm{m}_{\ell}+\bm{m}_{\ell+1})/|\bm{m}_{\ell}+\bm{m}_{\ell+1}|$ represents the unit vector of the uniform magnetization between the sublattices $\ell$ and $\ell+1$ (with $\bm{d}_{3,4}\equiv\bm{d}_{3,1}$). The BdG Hamiltonian satisfies the particle-hole symmetry $(\tau^{z}\otimes I)H_{\mathrm{BdG}}(\bm{k})(\tau^{z}\otimes I)=H_{\mathrm{BdG}}^{*}(-\bm{k})$ and can therefore be diagonalized by a paraunitary matrix $P(\bm{r})$ as~\cite{colpa1978pa}
\begin{align}
    P^{\dagger}(\bm{k})H_{\mathrm{BdG}}(\bm{k})P(\bm{k})=\mathrm{diag}\left(\varepsilon_{1}(\bm{k}),\varepsilon_{2}(\bm{k}),\varepsilon_{3}(\bm{k}),\varepsilon_{1}(-\bm{k}),\varepsilon_{2}(-\bm{k}),\varepsilon_{3}(-\bm{k})\right),
\end{align}
and $\varepsilon_{n}(\bm{k})$ ($\varepsilon_{1}(\bm{k})\leq\varepsilon_{2}(\bm{k})\leq\varepsilon_{3}(\bm{k})$) is the magnon band.
The paraunitary transformation of the bosonic operators $\hat{\Psi}_{\bm{k}}=P(\bm{k})\hat{\Psi}_{\bm{k}}'$ with $\hat{\Psi}_{\bm{k}}'=(\hat{\gamma}_{\bm{k},1},\hat{\gamma}_{\bm{k},2},\hat{\gamma}_{\bm{k},3},\hat{\gamma}_{-\bm{k},1}^{\dagger},\hat{\gamma}_{-\bm{k},2}^{\dagger},\hat{\gamma}_{-\bm{k},3}^{\dagger})^{T}$ leads to the diagonal form,
\begin{align}
    \hat{\mathcal{H}}\simeq\sum_{\bm{k}}\sum_{n=1}^{3}\varepsilon_{n}(\bm{k})\left(\hat{\gamma}_{\bm{k},n}^{\dagger}\hat{\gamma}_{\bm{k},n}+\frac{1}{2}\right).
\end{align}
The Berry curvature of the $n$th band ($n=1,2,3$) is calculated from $P(\bm{k})$ as~\cite{matsumoto2011prb,matsumoto2011prl,matsumoto2014prb}
\begin{align}
    \Omega_{xy}^{(n)}(\bm{k})=-2\mathrm{Im}\left[\frac{\partial P^{\dagger}(\bm{k})}{\partial k_{x}}(\tau^{z}\otimes I)\frac{\partial P(\bm{k})}{\partial k_{y}}\right]_{n,n}.
\end{align}

Figure~\ref{fig:application}(a) shows the magnon bands along high-symmetry lines in the Brillouin zone. The finite DM interaction lifts the degeneracy of the magnon bands near the symmetry points and generates the Berry curvature as shown in Fig.~\ref{fig:application}(b). In particular, the Berry curvature of the lowest band ($n=1$) is strongly enhanced around the $\Gamma$ point. This point is gapless, which may seem unusual because the DM and $\Lambda$ usually works to open a gap in the Heisenberg AFM. This is because the contribution from the DM interaction appears as the odd function of $\bm{k}$ in $H_{\mathrm{BdG}}(\bm{k})$ and therefore vanishes at the $\Gamma$ point. The $\Lambda$ term opens a gap for two bands among the three-fold degenerate bands and one gapless band remains.

Figure~\ref{fig:application}(c) shows the temperature dependence of the thermal Hall conductivity for several choices of 
the DM interaction $D$. 
Starting from the ground state, $\kappa_{xy}$ increases monotonically in magnitude with increasing temperature. 
This behavior comes from the contribution from the lowest magnon band, 
whose Berry curvature near the $\Gamma$ point has a positive sign in total. 
The rapid growth of $\kappa_{xy}$ with increasing temperature is because the magnon population grows rapidly around the 
gapless point of the lowest band. 

At higher temperatures, contributions from the higher-energy magnon bands also become relevant. 
In particular, the second band ($n=2$) carries the Berry curvature with the opposite sign, as shown in Fig.~\ref{fig:application}(b), 
and starts to suppress $\kappa_{xy}$ by cancelling part of the contributions from $n=1$. 
Then $\kappa_{xy}$ shows a maximum amplitude at around $k_BT\sim 0.4-0.6$ (in the unit of $JS=1.0$), 
which is the crossover between the single-band-dominated regime and the multiband regime.

Figure~\ref{fig:application}(d) shows the $D$ dependence of $\kappa_{xy}$. 
Although the field strength in Eq.~(\ref{eq:fxy-120-DM}) scales as $F_{xy}\propto D^{2}$ at small $D$, $\kappa_{xy}$ does not follow a simple quadratic dependence on $D$, 
but instead shows a sublinear growth. 
This behavior is in sharp contrast to kagome ferromagnets, where $\kappa_{xy}$ is typically proportional to the flux $\phi\propto D$~\cite{katsura2010prl}. 
Within the present framework, it is difficult to obtain a quantitative connection between 
$\kappa_{xy}$ and $F_{xy}$, which is left for a future study. 

\section{Conclusion}
In this work, we have systematically classified a broad class of standard 2D lattice spin models,
including ferromagnets, collinear and noncollinear antiferromagnets, and noncoplanar magnets including skyrmion crystals,
from the viewpoint of whether they can host a finite thermal Hall conductivity, $\kappa_{xy}\neq0$. 
The overall complexity of judging this issue lies in the diversity of crystalline symmetries and magnetic orders,
which makes a straightforward classification nontrivial.

At the microscopic level, lattice symmetries determine the types of interactions; 
the inversion symmetry constrains the allowed patterns of the DM interactions and other anisotropic couplings.
These constraints play a primary role in determining whether Berry curvature can be generated in magnon bands.
Beyond this, the compatibility between magnetic order and lattice geometry further dictates 
whether Berry curvatures cancel in reciprocal space, leading to $\kappa_{xy}=0$.
Such symmetry-enforced cancellations constitute what is referred to as the no-go rule for the thermal Hall effect.

We emphasized that the origin of Berry curvature in magnetic insulators 
cannot be fully captured by the conventional U(1) gauge-field picture developed for ferromagnets.
In particular, antiferromagnets with multiple magnetic sublattices naturally give rise to non-Abelian gauge structures,
closely analogous to the SU(2) gauge-field formulation of the electronic anomalous Hall effect 
originating from SOC.
Using a coarse-grained formulation, we showed that magnon systems with $N$ magnetic sublattices
are generically described by emergent SU($N$) gauge fields,
overcoming the apparent difficulty of making correspondence between the discrete sublattice degrees of freedom of magnons 
to the internal spin variables of electrons.

Within this framework, we first summarized known mechanisms that generate U(1) gauge fields in insulating ferromagnets
and clarified how symmetry lowering allows them to evade the no-go rule.
We then formulated the non-Abelian gauge theory for magnons and demonstrated that
the emergence of a finite SU($N$) gauge field generically guarantees a nonvanishing thermal Hall conductivity.
This observation constitutes a ``rule-to-go'':
unlike the Abelian case, noncommutativity of SU($N$) gauge fields leads to
path-dependent gauge fluxes for magnons moving along closed loops, 
naturally avoiding symmetry-enforced cancellations of Berry curvature.

Experimentally realized examples of non-Abelian-gauge induced magnon thermal Hall effects remain scarce.
At present, the AFM-SkX provides the only confirmed realization, 
where multiple ingredients, i.e., noncoplanarity, sublattice structure, and spatial modulation of spins, 
either cooperate or are partially chosen to generate an SU(3) gauge field. 
While this issue itself is solved by the field-theoretical treatment~\cite{kawano2025prb}, 
we need more general clues to understand how the microscopic magnetic structures or the interactions generate such gauge fields. 
To establish a simpler and more canonical platform, 
we finally presented a complete LSW analysis, 
demonstrating that the coplanar $120^{\circ}$ antiferromagnetic structure,
when combined with the DM interaction, realizes another concrete example of an SU(3)-gauge induced thermal Hall effect in magnon systems.

The comprehensive list of model systems and mechanisms summarized in this work provides 
a practical guideline for exploring thermal Hall effects in a wide variety of magnetic materials,
including recently proposed altermagnets~\cite{hayami2020prb,smejkal2022prx1,smejkal2022prx2}. 
A remaining issue on the antiferromagnetic thermal Hall effect is to make a direct connection with the 
electronic AHE.
For the square lattice Mott insulators, recent experiments on La-214, Nd$_{2}$CuO$_{4}$, and Sr$_{2}$CuO$_{2}$Cl$_{2}$ have reported
a remarkably large anomalous thermal Hall response, extending inside the pseudogap regime~\cite{grissonnanche2019nature,boulanger2020ncom}.
Related theoretical work shows that the thermal Hall effect arises intrinsically 
in a Mott insulating regime of electron systems
when both TRS and particle-hole symmetry are broken,
despite the absence of a finite Hall response within the LSW~\cite{ding-zhang2025prl}. 
Our earlier study on the insulating antiferromagnets on a square lattice~\cite{kawano2019prb1} 
can be regarded as their counterpart, using a reduced low-energy Hilbert space. 
Therefore, it would be interesting to clarify how the symmetries present in the original electronic systems are encoded 
or enlarged through successive approximations---from electronic models to effective spin Hamiltonian and further to magnon Hamiltonians---and conversely, how the symmetry of the original model is restored when higher-order superexchange processes
or magnon-magnon interactions are taken into account. 
\par
We finally note that phonon-assisted thermal Hall effects~\cite{strohm2005prl,qin2012prb}, including the so-called planar thermal Hall effect~\cite{chen2024prx,chen2024ncom,dhakal2024axv} and those arising from magnon-phonon hybridization~\cite{ideue2017nmat,zhang2019prl,park2020nl,akazawa2020prx,li2023prb,debnath2025prb2}, also constitute an important and experimentally accessible class of phenomena, 
for which the present framework shall be extended further in future studies. 

\section*{Acknowledgments}
This work is supported by a Grant-in-Aid for Transformative Research Areas ``The Natural Laws of Extreme Universe---A New Paradigm for Spacetime and Matter from Quantum Information'' (Grant No. 21H05191) and other JSPS KAKENHI (No. 21K03440).

\bibliographystyle{apsrev4-2}
\bibliography{biblio}

@article{holstein1940pr,
  title = {{Field Dependence of the Intrinsic Domain Magnetization of a Ferromagnet}},
  author = {Holstein, T. and Primakoff, H.},
  journal = {Phys. Rev.},
  volume = {58},
  issue = {12},
  pages = {1098--1113},
  numpages = {0},
  year = {1940},
  month = {Dec},
  publisher = {American Physical Society},
  doi = {10.1103/PhysRev.58.1098},
  url = {https://link.aps.org/doi/10.1103/PhysRev.58.1098}
}

@article{colpa1978pa,
  title = {{Diagonalization of the quadratic boson hamiltonian}},
  author = {J. H. P. Colpa},
  journal = {Physica A},
  volume = {93},
  pages = {327-353},
  year = {1978},
  url = {http://www.sciencedirect.com/science/article/pii/0378437178901607}
}

@article{onose2010science,
  title = {{Observation of the Magnon Hall Effect}},
  author = {Onose, Y. and Ideue, T. and Katsura, H. and Shiomi, Y. and Nagaosa, N. and Tokura, Y.},
  volume = {329},
  number = {5989},
  pages = {297--299},
  year = {2010},
  doi = {10.1126/science.1188260},
  publisher = {American Association for the Advancement of Science},
  journal = {Science}
}

@article{ideue2012prb,
  title = {{Effect of lattice geometry on magnon Hall effect in ferromagnetic insulators}},
  author = {Ideue, T. and Onose, Y. and Katsura, H. and Shiomi, Y. and Ishiwata, S. and Nagaosa, N. and Tokura, Y.},
  journal = {Phys. Rev. B},
  volume = {85},
  issue = {13},
  pages = {134411},
  numpages = {11},
  year = {2012},
  month = {Apr},
  publisher = {American Physical Society},
  doi = {10.1103/PhysRevB.85.134411},
  url = {https://link.aps.org/doi/10.1103/PhysRevB.85.134411}
}

@article{hirschberger2015prl,
  title = {{Thermal Hall Effect of Spin Excitations in a Kagome Magnet}},
  author = {Hirschberger, Max and Chisnell, Robin and Lee, Young S. and Ong, N. P.},
  journal = {Phys. Rev. Lett.},
  volume = {115},
  issue = {10},
  pages = {106603},
  numpages = {5},
  year = {2015},
  month = {Sep},
  publisher = {American Physical Society},
  doi = {10.1103/PhysRevLett.115.106603},
  url = {https://link.aps.org/doi/10.1103/PhysRevLett.115.106603}
}

@article{akazawa2022prr,
  title = {{Topological thermal Hall effect of magnons in magnetic skyrmion lattice}},
  author = {Akazawa, Masatoshi and Lee, Hyun-Yong and Takeda, Hikaru and Fujima, Yuri and Tokunaga, Yusuke and Arima, Taka-hisa and Han, Jung Hoon and Yamashita, Minoru},
  journal = {Phys. Rev. Res.},
  volume = {4},
  issue = {4},
  pages = {043085},
  numpages = {16},
  year = {2022},
  month = {Nov},
  publisher = {American Physical Society},
  doi = {10.1103/PhysRevResearch.4.043085},
  url = {https://link.aps.org/doi/10.1103/PhysRevResearch.4.043085}
}

@article{fujimoto2009prl,
  title = {{Hall Effect of Spin Waves in Frustrated Magnets}},
  author = {Fujimoto, Satoshi},
  journal = {Phys. Rev. Lett.},
  volume = {103},
  issue = {4},
  pages = {047203},
  numpages = {4},
  year = {2009},
  month = {Jul},
  publisher = {American Physical Society},
  doi = {10.1103/PhysRevLett.103.047203},
  url = {https://link.aps.org/doi/10.1103/PhysRevLett.103.047203}
}

@article{katsura2010prl,
  title = {{Theory of the Thermal Hall Effect in Quantum Magnets}},
  author = {Katsura, Hosho and Nagaosa, Naoto and Lee, Patrick A.},
  journal = {Phys. Rev. Lett.},
  volume = {104},
  issue = {6},
  pages = {066403},
  numpages = {4},
  year = {2010},
  month = {Feb},
  publisher = {American Physical Society},
  doi = {10.1103/PhysRevLett.104.066403},
  url = {https://link.aps.org/doi/10.1103/PhysRevLett.104.066403}
}

@article{matsumoto2011prl,
  title = {{Theoretical Prediction of a Rotating Magnon Wave Packet in Ferromagnets}},
  author = {Matsumoto, Ryo and Murakami, Shuichi},
  journal = {Phys. Rev. Lett.},
  volume = {106},
  issue = {19},
  pages = {197202},
  numpages = {4},
  year = {2011},
  month = {May},
  publisher = {American Physical Society},
  doi = {10.1103/PhysRevLett.106.197202},
  url = {https://link.aps.org/doi/10.1103/PhysRevLett.106.197202}
}

@article{matsumoto2011prb,
  title = {{Rotational motion of magnons and the thermal Hall effect}},
  author = {Matsumoto, Ryo and Murakami, Shuichi},
  journal = {Phys. Rev. B},
  volume = {84},
  issue = {18},
  pages = {184406},
  numpages = {9},
  year = {2011},
  month = {Nov},
  publisher = {American Physical Society},
  doi = {10.1103/PhysRevB.84.184406},
  url = {https://link.aps.org/doi/10.1103/PhysRevB.84.184406}
}

@article{matsumoto2014prb,
  title = {{Thermal Hall effect of magnons in magnets with dipolar interaction}},
  author = {Matsumoto, Ryo and Shindou, Ryuichi and Murakami, Shuichi},
  journal = {Phys. Rev. B},
  volume = {89},
  issue = {5},
  pages = {054420},
  numpages = {12},
  year = {2014},
  month = {Feb},
  publisher = {American Physical Society},
  doi = {10.1103/PhysRevB.89.054420},
  url = {https://link.aps.org/doi/10.1103/PhysRevB.89.054420}
}

@article{mook2014prb1,
  title = {{Edge states in topological magnon insulators}},
  author = {Mook, Alexander and Henk, J\"urgen and Mertig, Ingrid},
  journal = {Phys. Rev. B},
  volume = {90},
  issue = {2},
  pages = {024412},
  numpages = {7},
  year = {2014},
  month = {Jul},
  publisher = {American Physical Society},
  doi = {10.1103/PhysRevB.90.024412},
  url = {https://link.aps.org/doi/10.1103/PhysRevB.90.024412}
}

@article{mook2014prb2,
  title = {{Magnon Hall effect and topology in kagome lattices: A theoretical investigation}},
  author = {Mook, Alexander and Henk, J\"urgen and Mertig, Ingrid},
  journal = {Phys. Rev. B},
  volume = {89},
  issue = {13},
  pages = {134409},
  numpages = {9},
  year = {2014},
  month = {Apr},
  publisher = {American Physical Society},
  doi = {10.1103/PhysRevB.89.134409},
  url = {https://link.aps.org/doi/10.1103/PhysRevB.89.134409}
}

@article{mook2016prb,
  title = {{Spin dynamics simulations of topological magnon insulators: From transverse current correlation functions to the family of magnon Hall effects}},
  author = {Mook, Alexander and Henk, J\"urgen and Mertig, Ingrid},
  journal = {Phys. Rev. B},
  volume = {94},
  issue = {17},
  pages = {174444},
  numpages = {13},
  year = {2016},
  month = {Nov},
  publisher = {American Physical Society},
  doi = {10.1103/PhysRevB.94.174444},
  url = {https://link.aps.org/doi/10.1103/PhysRevB.94.174444}
}

@article{owerre2017prb,
  title = {{Topological thermal Hall effect in frustrated kagome antiferromagnets}},
  author = {Owerre, S. A.},
  journal = {Phys. Rev. B},
  volume = {95},
  issue = {1},
  pages = {014422},
  numpages = {6},
  year = {2017},
  month = {Jan},
  publisher = {American Physical Society},
  doi = {10.1103/PhysRevB.95.014422},
  url = {https://link.aps.org/doi/10.1103/PhysRevB.95.014422}
}

@article{owerre2018prb,
  title = {{Weyl magnons in noncoplanar stacked kagome antiferromagnets}},
  author = {Owerre, S. A.},
  journal = {Phys. Rev. B},
  volume = {97},
  issue = {9},
  pages = {094412},
  numpages = {11},
  year = {2018},
  month = {Mar},
  publisher = {American Physical Society},
  doi = {10.1103/PhysRevB.97.094412},
  url = {https://link.aps.org/doi/10.1103/PhysRevB.97.094412}
}

@article{seshadri2018prb,
  title = {{Topological magnons in a kagome-lattice spin system with $XXZ$ and Dzyaloshinskii-Moriya interactions}},
  author = {Seshadri, Ranjani and Sen, Diptiman},
  journal = {Phys. Rev. B},
  volume = {97},
  issue = {13},
  pages = {134411},
  numpages = {14},
  year = {2018},
  month = {Apr},
  publisher = {American Physical Society},
  doi = {10.1103/PhysRevB.97.134411},
  url = {https://link.aps.org/doi/10.1103/PhysRevB.97.134411}
}

@article{laurell2018prb,
  title = {{Magnon thermal Hall effect in kagome antiferromagnets with Dzyaloshinskii-Moriya interactions}},
  author = {Laurell, Pontus and Fiete, Gregory A.},
  journal = {Phys. Rev. B},
  volume = {98},
  issue = {9},
  pages = {094419},
  numpages = {10},
  year = {2018},
  month = {Sep},
  publisher = {American Physical Society},
  doi = {10.1103/PhysRevB.98.094419},
  url = {https://link.aps.org/doi/10.1103/PhysRevB.98.094419}
}

@article{owerre2017epl,
  title={{Topological magnetic excitations on the distorted kagom{\'e} antiferromagnets: Applications to volborthite, vesignieite, and edwardsite}},
  author={Owerre, SA},
  journal={Europhys. Lett.},
  volume={117},
  number={3},
  pages={37006},
  year={2017},
  publisher={IOP Publishing},
  url={https://iopscience.iop.org/article/10.1209/0295-5075/117/37006}
}

@article{owerre2018sp,
  title={{Photoinduced topological phase transitions in topological magnon insulators}},
  author={Owerre, SA},
  journal={Sci. Rep.},
  volume={8},
  number={1},
  pages={4431},
  year={2018},
  publisher={Nature Publishing Group UK London},
  utl={https://www.nature.com/articles/s41598-018-22779-8}
}

@article{owerre2019ap,
  title={{Photo-induced Floquet Weyl magnons in noncollinear antiferromagnets}},
  author={Owerre, SA},
  journal={Ann. Phys.},
  volume={406},
  pages={14--29},
  year={2019},
  publisher={Elsevier},
  url={https://www.sciencedirect.com/science/article/pii/S0003491619300934?via%3Dihub}
}

@article{laurell2017prl,
  title = {{Topological Magnon Bands and Unconventional Superconductivity in Pyrochlore Iridate Thin Films}},
  author = {Laurell, Pontus and Fiete, Gregory A.},
  journal = {Phys. Rev. Lett.},
  volume = {118},
  issue = {17},
  pages = {177201},
  numpages = {7},
  year = {2017},
  month = {Apr},
  publisher = {American Physical Society},
  doi = {10.1103/PhysRevLett.118.177201},
  url = {https://link.aps.org/doi/10.1103/PhysRevLett.118.177201}
}

@article{kim2019prb,
  title = {{Magnon topology and thermal Hall effect in trimerized triangular lattice antiferromagnet}},
  author = {Kim, Kyung-Su and Lee, Ki Hoon and Chung, Suk Bum and Park, Je-Geun},
  journal = {Phys. Rev. B},
  volume = {100},
  issue = {6},
  pages = {064412},
  numpages = {6},
  year = {2019},
  month = {Aug},
  publisher = {American Physical Society},
  doi = {10.1103/PhysRevB.100.064412},
  url = {https://link.aps.org/doi/10.1103/PhysRevB.100.064412}
}

@article{kim2024ncom,
  title = {{Thermal Hall effects due to topological spin fluctuations in YMnO3}},
  author = {Kim, Ha-Leem and Saito, Takuma and Yang, Heejun and Ishizuka, Hiroaki and Coak, Matthew John and Lee, Jun Han and Sim, Hasung and Oh, Yoon Seok and Nagaosa, Naoto and Park, Je-Geun},
  journal = {Nat. Commun.},
  volume = {15},
  number = {1},
  pages = {243},
  year = {2024},
  publisher = {Nature Publishing Group UK London},
  doi = {10.1038/s41467-023-44448-9},
  url = {https://doi.org/10.1038/s41467-023-44448-9}
}

@article{owerre2016jap,
  author = {Owerre,S. A. },
  title = {{Topological honeycomb magnon Hall effect: A calculation of thermal Hall conductivity of magnetic spin excitations}},
  journal = {J. Appl. Phys.},
  volume = {120},
  number = {4},
  pages = {043903},
  year = {2016},
  doi = {10.1063/1.4959815},
  url = {https://doi.org/10.1063/1.4959815}
}

@article{owerre2016prb,
  title={{Magnon Hall effect in AB-stacked bilayer honeycomb quantum magnets}},
  author={Owerre, SA},
  journal={Phys. Rev. B},
  volume={94},
  number={9},
  pages={094405},
  year={2016},
  publisher={APS}
}

@article{owerre2017jpcm,
  title={{Topological magnon bands and unconventional thermal Hall effect on the frustrated honeycomb and bilayer triangular lattice}},
  author={Owerre, SA},
  journal={J. Phys.: Cond. Matt.},
  volume={29},
  number={38},
  pages={385801},
  year={2017},
  publisher={IOP Publishing},
  url={https://iopscience.iop.org/article/10.1088/1361-648X/aa7dd2}
}

@article{owerre2017jap,
  title={{Noncollinear antiferromagnetic Haldane magnon insulator}},
  author={Owerre, SA},
  journal={J. Appl. Phys.},
  volume={121},
  number={22},
  pages={223904},
  year={2017},
  publisher={AIP Publishing}
}

@article{lu2021prl,
  title = {{Topological Phase Transitions of Dirac Magnons in Honeycomb Ferromagnets}},
  author = {Lu, Yu-Shan and Li, Jian-Lin and Wu, Chien-Te},
  journal = {Phys. Rev. Lett.},
  volume = {127},
  issue = {21},
  pages = {217202},
  numpages = {6},
  year = {2021},
  month = {Nov},
  publisher = {American Physical Society},
  doi = {10.1103/PhysRevLett.127.217202},
  url = {https://link.aps.org/doi/10.1103/PhysRevLett.127.217202}
}

@article{mcclarty2018prb,
  title = {{Topological magnons in Kitaev magnets at high fields}},
  author = {McClarty, P. A. and Dong, X.-Y. and Gohlke, M. and Rau, J. G. and Pollmann, F. and Moessner, R. and Penc, K.},
  journal = {Phys. Rev. B},
  volume = {98},
  issue = {6},
  pages = {060404},
  numpages = {6},
  year = {2018},
  month = {Aug},
  publisher = {American Physical Society},
  doi = {10.1103/PhysRevB.98.060404},
  url = {https://link.aps.org/doi/10.1103/PhysRevB.98.060404}
}

@article{neumann2022prl,
  title = {{Thermal Hall Effect of Magnons in Collinear Antiferromagnetic Insulators: Signatures of Magnetic and Topological Phase Transitions}},
  author = {Neumann, Robin R. and Mook, Alexander and Henk, J\"urgen and Mertig, Ingrid},
  journal = {Phys. Rev. Lett.},
  volume = {128},
  issue = {11},
  pages = {117201},
  numpages = {8},
  year = {2022},
  month = {Mar},
  publisher = {American Physical Society},
  doi = {10.1103/PhysRevLett.128.117201},
  url = {https://link.aps.org/doi/10.1103/PhysRevLett.128.117201}
}

@article{chern2021nqm,
  title={{Classical magnetic vortex liquid and large thermal Hall conductivity in frustrated magnets with bond-dependent interactions}},
  author={Chern, Li Ern and Buessen, Finn Lasse and Kim, Yong Baek},
  journal={npj Quantum Materials},
  volume={6},
  number={1},
  pages={33},
  year={2021},
  publisher={Nature Publishing Group UK London},
  doi = {10.1038/s41535-021-00331-8},
  url = {https://doi.org/10.1038/s41535-021-00331-8}
}

@article{koyama2021prb,
  title = {{Field-angle dependence of thermal Hall conductivity in a magnetically ordered Kitaev-Heisenberg system}},
  author = {Koyama, Shinnosuke and Nasu, Joji},
  journal = {Phys. Rev. B},
  volume = {104},
  issue = {7},
  pages = {075121},
  numpages = {13},
  year = {2021},
  month = {Aug},
  publisher = {American Physical Society},
  doi = {10.1103/PhysRevB.104.075121},
  url = {https://link.aps.org/doi/10.1103/PhysRevB.104.075121}
}

@article{chern2020prr,
  title = {{Magnetic field induced competing phases in spin-orbital entangled Kitaev magnets}},
  author = {Chern, Li Ern and Kaneko, Ryui and Lee, Hyun-Yong and Kim, Yong Baek},
  journal = {Phys. Rev. Res.},
  volume = {2},
  issue = {1},
  pages = {013014},
  numpages = {9},
  year = {2020},
  month = {Jan},
  publisher = {American Physical Society},
  doi = {10.1103/PhysRevResearch.2.013014},
  url = {https://link.aps.org/doi/10.1103/PhysRevResearch.2.013014}
}

@article{zhang2021prb,
  title = {{Topological magnons for thermal Hall transport in frustrated magnets with bond-dependent interactions}},
  author = {Zhang, Emily Z. and Chern, Li Ern and Kim, Yong Baek},
  journal = {Phys. Rev. B},
  volume = {103},
  issue = {17},
  pages = {174402},
  numpages = {9},
  year = {2021},
  month = {May},
  publisher = {American Physical Society},
  doi = {10.1103/PhysRevB.103.174402},
  url = {https://link.aps.org/doi/10.1103/PhysRevB.103.174402}
}

@article{chern2021prl,
  title = {{Sign Structure of Thermal Hall Conductivity and Topological Magnons for In-Plane Field Polarized Kitaev Magnets}},
  author = {Chern, Li Ern and Zhang, Emily Z. and Kim, Yong Baek},
  journal = {Phys. Rev. Lett.},
  volume = {126},
  issue = {14},
  pages = {147201},
  numpages = {6},
  year = {2021},
  month = {Apr},
  publisher = {American Physical Society},
  doi = {10.1103/PhysRevLett.126.147201},
  url = {https://link.aps.org/doi/10.1103/PhysRevLett.126.147201}
}

@article{nakata2017prb1,
  title = {{Magnonic quantum Hall effect and Wiedemann-Franz law}},
  author = {Nakata, Kouki and Klinovaja, Jelena and Loss, Daniel},
  journal = {Phys. Rev. B},
  volume = {95},
  issue = {12},
  pages = {125429},
  numpages = {11},
  year = {2017},
  month = {Mar},
  publisher = {American Physical Society},
  doi = {10.1103/PhysRevB.95.125429},
  url = {https://link.aps.org/doi/10.1103/PhysRevB.95.125429}
}

@article{nakata2017prb2,
  title = {{Magnonic topological insulators in antiferromagnets}},
  author = {Nakata, Kouki and Kim, Se Kwon and Klinovaja, Jelena and Loss, Daniel},
  journal = {Phys. Rev. B},
  volume = {96},
  issue = {22},
  pages = {224414},
  numpages = {14},
  year = {2017},
  month = {Dec},
  publisher = {American Physical Society},
  doi = {10.1103/PhysRevB.96.224414},
  url = {https://link.aps.org/doi/10.1103/PhysRevB.96.224414}
}

@article{hoogdalem2013prb,
  title = {{Magnetic texture-induced thermal Hall effects}},
  author = {van Hoogdalem, Kevin A. and Tserkovnyak, Yaroslav and Loss, Daniel},
  journal = {Phys. Rev. B},
  volume = {87},
  issue = {2},
  pages = {024402},
  numpages = {7},
  year = {2013},
  month = {Jan},
  publisher = {American Physical Society},
  doi = {10.1103/PhysRevB.87.024402},
  url = {https://link.aps.org/doi/10.1103/PhysRevB.87.024402}
}

@article{kong2013prl,
  title = {{Dynamics of an Insulating Skyrmion under a Temperature Gradient}},
  author = {Kong, Lingyao and Zang, Jiadong},
  journal = {Phys. Rev. Lett.},
  volume = {111},
  issue = {6},
  pages = {067203},
  numpages = {5},
  year = {2013},
  month = {Aug},
  publisher = {American Physical Society},
  doi = {10.1103/PhysRevLett.111.067203},
  url = {https://link.aps.org/doi/10.1103/PhysRevLett.111.067203}
}

@article{iwasaki2014prb,
  title = {{Theory of magnon-skyrmion scattering in chiral magnets}},
  author = {Iwasaki, Junichi and Beekman, Aron J. and Nagaosa, Naoto},
  journal = {Phys. Rev. B},
  volume = {89},
  issue = {6},
  pages = {064412},
  numpages = {7},
  year = {2014},
  month = {Feb},
  publisher = {American Physical Society},
  doi = {10.1103/PhysRevB.89.064412},
  url = {https://link.aps.org/doi/10.1103/PhysRevB.89.064412}
}

@article{oh2015prb,
  title = {{Dynamics of magnon fluid in Dzyaloshinskii-Moriya magnet and its manifestation in magnon-Skyrmion scattering}},
  author = {Oh, Yun-Tak and Lee, Hyunyong and Park, Jin-Hong and Han, Jung Hoon},
  journal = {Phys. Rev. B},
  volume = {91},
  issue = {10},
  pages = {104435},
  numpages = {6},
  year = {2015},
  month = {Mar},
  publisher = {American Physical Society},
  doi = {10.1103/PhysRevB.91.104435},
  url = {https://link.aps.org/doi/10.1103/PhysRevB.91.104435}
}

@article{roldan2016njp,
  title = {{Topological spin waves in the atomic-scale magnetic skyrmion crystal}},
  author = {A Roldán-Molina and A S Nunez and J Fernández-Rossier},
  journal = {New J. Phys.},
  volume = {18},
  pages = {045015},
  year = {2016},
  doi = {10.1088/1367-2630/18/4/045015},
  url = {https://dx.doi.org/10.1088/1367-2630/18/4/045015}
}

@article{mook2017prb,
  title = {{Magnon transport in noncollinear spin textures: Anisotropies and topological magnon Hall effects}},
  author = {Mook, Alexander and G\"obel, B\"orge and Henk, J\"urgen and Mertig, Ingrid},
  journal = {Phys. Rev. B},
  volume = {95},
  issue = {2},
  pages = {020401},
  numpages = {5},
  year = {2017},
  month = {Jan},
  publisher = {American Physical Society},
  doi = {10.1103/PhysRevB.95.020401},
  url = {https://link.aps.org/doi/10.1103/PhysRevB.95.020401}
}

@article{kim2019prl,
  title = {{Tunable Magnonic Thermal Hall Effect in Skyrmion Crystal Phases of Ferrimagnets}},
  author = {Kim, Se Kwon and Nakata, Kouki and Loss, Daniel and Tserkovnyak, Yaroslav},
  journal = {Phys. Rev. Lett.},
  volume = {122},
  issue = {5},
  pages = {057204},
  numpages = {6},
  year = {2019},
  month = {Feb},
  publisher = {American Physical Society},
  doi = {10.1103/PhysRevLett.122.057204},
  url = {https://link.aps.org/doi/10.1103/PhysRevLett.122.057204}
}

@article{nikolic2020prb,
  title = {{Quantum field theory of topological spin dynamics}},
  author = {Nikoli\ifmmode \acute{c}\else \'{c}\fi{}, Predrag},
  journal = {Phys. Rev. B},
  volume = {102},
  issue = {7},
  pages = {075131},
  numpages = {40},
  year = {2020},
  month = {Aug},
  publisher = {American Physical Society},
  doi = {10.1103/PhysRevB.102.075131},
  url = {https://link.aps.org/doi/10.1103/PhysRevB.102.075131}
}

@article{cao2015jpcm,
  author={Xiaodong Cao and Kai Chen and Dahai He},
  title={{Magnon Hall effect on the Lieb lattice}},
  journal={J. Phys: Condens. Matter},
  volume={27},
  number={16},
  pages={166003},
  url={http://stacks.iop.org/0953-8984/27/i=16/a=166003},
  year={2015},
}

@article{owerre2016jpcm,
  title = {{Magnon Hall effect without Dzyaloshinskii–Moriya interaction}},
  author = {Owerre, S A},
  journal = {J. Phys.: Cond. Matt},
  volume = {29},
  pages = {03LT01},
  year = {2016},
  publisher={IOP Publishing},
  url = {https://doi.org/10.1088/0953-8984/29/3/03LT01}
}

@article{judit2015ncom,
  title = {{Hall effect of triplons in a dimerized quantum magnet}},
  author = {Romh\'anyi, Judit and Penc, Karlo and Ganesh, R},
  journal = {Nat. Commun.},
  volume = {6},
  number = {6805},
  year = {2015},
  publisher = {Nature Publishing Group, a division of Macmillan Publishers Limited. All Rights Reserved.},
  doi = {10.1038/ncomms7805},
  url = {https://doi.org/10.1038/ncomms7805}
}

@article{malki2019prb,
  title = {{Topological magnon bands for magnonics}},
  author = {Malki, M. and Uhrig, G. S.},
  journal = {Phys. Rev. B},
  volume = {99},
  issue = {17},
  pages = {174412},
  numpages = {5},
  year = {2019},
  month = {May},
  publisher = {American Physical Society},
  doi = {10.1103/PhysRevB.99.174412},
  url = {https://link.aps.org/doi/10.1103/PhysRevB.99.174412}
}

@article{buzo2024prb,
  title = {{Thermal Hall conductivity of a valence bond solid phase in the square lattice ${J}_{1}\text{\ensuremath{-}}{J}_{2}$ antiferromagnet Heisenberg model with a Dzyaloshinskii-Moriya interaction}},
  author = {Buzo, Lucas S. and Doretto, R. L.},
  journal = {Phys. Rev. B},
  volume = {109},
  issue = {13},
  pages = {134405},
  numpages = {18},
  year = {2024},
  month = {Apr},
  publisher = {American Physical Society},
  doi = {10.1103/PhysRevB.109.134405},
  url = {https://link.aps.org/doi/10.1103/PhysRevB.109.134405}
}

@article{suetsugu2022prb,
  title = {{Intrinsic suppression of the topological thermal Hall effect in an exactly solvable quantum magnet}},
  author = {Suetsugu, S. and Yokoi, T. and Totsuka, K. and Ono, T. and Tanaka, I. and Kasahara, S. and Kasahara, Y. and Chengchao, Z. and Kageyama, H. and Matsuda, Y.},
  journal = {Phys. Rev. B},
  volume = {105},
  issue = {2},
  pages = {024415},
  numpages = {11},
  year = {2022},
  month = {Jan},
  publisher = {American Physical Society},
  doi = {10.1103/PhysRevB.105.024415},
  url = {https://link.aps.org/doi/10.1103/PhysRevB.105.024415}
}

@article{hoyer2025prb,
  title = {{Spontaneous crystal thermal Hall effect in insulating altermagnets}},
  author = {Hoyer, Rhea and Jaeschke-Ubiergo, Rodrigo and Ahn, Kyo-Hoon and \ifmmode \check{S}\else \v{S}\fi{}mejkal, Libor and Mook, Alexander},
  journal = {Phys. Rev. B},
  volume = {111},
  issue = {2},
  pages = {L020412},
  numpages = {8},
  year = {2025},
  month = {Jan},
  publisher = {American Physical Society},
  doi = {10.1103/PhysRevB.111.L020412},
  url = {https://link.aps.org/doi/10.1103/PhysRevB.111.L020412}
}

@article{dzyaloshinskii1958jpcs,
  title = {{A thermodynamic theory of ``weak'' ferromagnetism of antiferromagnetics}},
  journal = {J. Phys. Chem. Solids},
  volume = {4},
  number = {4},
  pages = {241 - 255},
  year = {1958},
  issn = {0022-3697},
  doi = {10.1016/0022-3697(58)90076-3},
  url = {http://www.sciencedirect.com/science/article/pii/0022369758900763},
  author = {I. Dzyaloshinsky},
}

@article{moriya1960pr,
  title = {{Anisotropic Superexchange Interaction and Weak Ferromagnetism}},
  author = {Moriya, T\^oru},
  journal = {Phys. Rev.},
  volume = {120},
  issue = {1},
  pages = {91--98},
  numpages = {0},
  year = {1960},
  month = {Oct},
  publisher = {American Physical Society},
  doi = {10.1103/PhysRev.120.91},
  url = {https://link.aps.org/doi/10.1103/PhysRev.120.91}
}

@article{wen1989prb,
  title = {{Chiral spin states and superconductivity}},
  author = {Wen, X. G. and Wilczek, Frank and Zee, A.},
  journal = {Phys. Rev. B},
  volume = {39},
  issue = {16},
  pages = {11413--11423},
  numpages = {0},
  year = {1989},
  month = {Jun},
  publisher = {American Physical Society},
  doi = {10.1103/PhysRevB.39.11413},
  url = {https://link.aps.org/doi/10.1103/PhysRevB.39.11413}
}

@article{ye1999prl,
  title = {{Berry Phase Theory of the Anomalous Hall Effect: Application to Colossal Magnetoresistance Manganites}},
  author = {Ye, Jinwu and Kim, Yong Baek and Millis, A. J. and Shraiman, B. I. and Majumdar, P. and Te\ifmmode \check{s}\else \v{s}\fi{}anovi\ifmmode \acute{c}\else \'{c}\fi{}, Z.},
  journal = {Phys. Rev. Lett.},
  volume = {83},
  issue = {18},
  pages = {3737--3740},
  numpages = {0},
  year = {1999},
  month = {Nov},
  publisher = {American Physical Society},
  doi = {10.1103/PhysRevLett.83.3737},
  url = {https://link.aps.org/doi/10.1103/PhysRevLett.83.3737}
}

@article{sen1995prb,
  title = {{Large-U limit of a Hubbard model in a magnetic field: Chiral spin interactions and paramagnetism}},
  author = {Sen, Diptiman and Chitra, R.},
  journal = {Phys. Rev. B},
  volume = {51},
  issue = {3},
  pages = {1922--1925},
  numpages = {0},
  year = {1995},
  month = {Jan},
  publisher = {American Physical Society},
  doi = {10.1103/PhysRevB.51.1922},
  url = {https://link.aps.org/doi/10.1103/PhysRevB.51.1922}
}

@article{kitaev2006ap,
  title = {{Anyons in an exactly solved model and beyond}},
  author = {Alexei Kitaev},
  journal = {Annals of Physics},
  volume = {321},
  issue = {1},
  pages = {2-111},
  year = {2006},
  doi = {10.1016/j.aop.2005.10.005}
}

@article{jackeli2009prl,
  title = {{Mott Insulators in the Strong Spin-Orbit Coupling Limit: From Heisenberg to a Quantum Compass and Kitaev Models}},
  author = {Jackeli, G. and Khaliullin, G.},
  journal = {Phys. Rev. Lett.},
  volume = {102},
  issue = {1},
  pages = {017205},
  numpages = {4},
  year = {2009},
  month = {Jan},
  publisher = {American Physical Society},
  doi = {10.1103/PhysRevLett.102.017205},
  url = {https://link.aps.org/doi/10.1103/PhysRevLett.102.017205}
}

@article{chaloupka2010prl,
  title = {{Kitaev-Heisenberg Model on a Honeycomb Lattice: Possible Exotic Phases in Iridium Oxides ${A}_{2}{\mathrm{IrO}}_{3}$}},
  author = {Chaloupka, Ji and Jackeli, George and Khaliullin, Giniyat},
  journal = {Phys. Rev. Lett.},
  volume = {105},
  issue = {2},
  pages = {027204},
  numpages = {4},
  year = {2010},
  month = {Jul},
  publisher = {American Physical Society},
  doi = {10.1103/PhysRevLett.105.027204},
  url = {https://link.aps.org/doi/10.1103/PhysRevLett.105.027204}
}

@article{rau2014prl,
  title = {{Generic Spin Model for the Honeycomb Iridates beyond the Kitaev Limit}},
  author = {Rau, Jeffrey G. and Lee, Eric Kin-Ho and Kee, Hae-Young},
  journal = {Phys. Rev. Lett.},
  volume = {112},
  issue = {7},
  pages = {077204},
  numpages = {5},
  year = {2014},
  month = {Feb},
  publisher = {American Physical Society},
  doi = {10.1103/PhysRevLett.112.077204},
  url = {https://link.aps.org/doi/10.1103/PhysRevLett.112.077204}
}

@article{mook2021prx,
  title = {{Interaction-Stabilized Topological Magnon Insulator in Ferromagnets}},
  author = {Mook, Alexander and Plekhanov, Kirill and Klinovaja, Jelena and Loss, Daniel},
  journal = {Phys. Rev. X},
  volume = {11},
  issue = {2},
  pages = {021061},
  numpages = {30},
  year = {2021},
  month = {Jun},
  publisher = {American Physical Society},
  doi = {10.1103/PhysRevX.11.021061},
  url = {https://link.aps.org/doi/10.1103/PhysRevX.11.021061}
}

@article{chatzichrysafis2025prb,
  title = {{Thermal Hall effect of magnons from many-body skew scattering}},
  author = {Chatzichrysafis, Dimos and Mook, Alexander},
  journal = {Phys. Rev. B},
  volume = {111},
  issue = {13},
  pages = {134405},
  numpages = {42},
  year = {2025},
  month = {Apr},
  publisher = {American Physical Society},
  doi = {10.1103/PhysRevB.111.134405},
  url = {https://link.aps.org/doi/10.1103/PhysRevB.111.134405}
}

@article{rosales2015prb,
  title = {{Three-sublattice skyrmion crystal in the antiferromagnetic triangular lattice}},
  author = {Rosales, H. D. and Cabra, D. C. and Pujol, Pierre},
  journal = {Phys. Rev. B},
  volume = {92},
  issue = {21},
  pages = {214439},
  numpages = {7},
  year = {2015},
  month = {Dec},
  publisher = {American Physical Society},
  doi = {10.1103/PhysRevB.92.214439},
  url = {https://link.aps.org/doi/10.1103/PhysRevB.92.214439}
}

@article{diaz2019prl,
  title = {{Topological Magnons and Edge States in Antiferromagnetic Skyrmion Crystals}},
  author = {D\'{\i}az, Sebasti\'an A. and Klinovaja, Jelena and Loss, Daniel},
  journal = {Phys. Rev. Lett.},
  volume = {122},
  issue = {18},
  pages = {187203},
  numpages = {6},
  year = {2019},
  month = {May},
  publisher = {American Physical Society},
  doi = {10.1103/PhysRevLett.122.187203},
  url = {https://link.aps.org/doi/10.1103/PhysRevLett.122.187203}
}

@article{mukherjee2021scirep,
  title = {{Antiferromagnetic skyrmion crystals in the Rashba Hund’s insulator on triangular lattice}},
  author = {Mukherjee, Arnob and Kathyat, Deepak S. and Kumar, Sanjeev},
  journal = {Sci. Rep.},
  volume = {11},
  pages = {9566},
  year = {2021},
  doi = {10.1038/s41598-021-88556-2},
  url = {https://doi.org/10.1038/s41598-021-88556-2}
}

@article{mohylna2022prb,
  title = {{Spontaneous antiferromagnetic skyrmion/antiskyrmion lattice and spiral spin-liquid states in the frustrated triangular lattice}},
  author = {Mohylna, M. and G\'omez Albarrac\'{\i}n, F. A. and \ifmmode \check{Z}\else \v{Z}\fi{}ukovi\ifmmode \check{c}\else \v{c}\fi{}, M. and Rosales, H. D.},
  journal = {Phys. Rev. B},
  volume = {106},
  issue = {22},
  pages = {224406},
  numpages = {10},
  year = {2022},
  month = {Dec},
  publisher = {American Physical Society},
  doi = {10.1103/PhysRevB.106.224406},
  url = {https://link.aps.org/doi/10.1103/PhysRevB.106.224406}
}

@article{kawano2019comphys,
  title = {{Designing Rashba--Dresselhaus effect in magnetic insulators}},
  author = {Kawano, Masataka and Onose, Yoshinori and Hotta, Chisa},
  journal = {Commun. Phys.},
  volume = {2},
  number = {1},
  pages = {27},
  year = {2019},
  publisher = {Nature Publishing Group UK London},
  doi = {10.1038/s42005-019-0128-6},
  url = {https://doi.org/10.1038/s42005-019-0128-6}
}

@article{kawano2019prb1,
  title = {{Thermal Hall effect and topological edge states in a square-lattice antiferromagnet}},
  author = {Kawano, Masataka and Hotta, Chisa},
  journal = {Phys. Rev. B},
  volume = {99},
  issue = {5},
  pages = {054422},
  numpages = {16},
  year = {2019},
  month = {Feb},
  publisher = {American Physical Society},
  doi = {10.1103/PhysRevB.99.054422},
  url = {https://link.aps.org/doi/10.1103/PhysRevB.99.054422}
}

@article{kawano2019prb2,
  title = {{Discovering momentum-dependent magnon spin texture in insulating antiferromagnets: Role of the Kitaev interaction}},
  author = {Kawano, Masataka and Hotta, Chisa},
  journal = {Phys. Rev. B},
  volume = {100},
  issue = {17},
  pages = {174402},
  numpages = {17},
  year = {2019},
  month = {Nov},
  publisher = {American Physical Society},
  doi = {10.1103/PhysRevB.100.174402},
  url = {https://link.aps.org/doi/10.1103/PhysRevB.100.174402}
}

@article{takeda2024ncom,
  title = {{Magnon thermal Hall effect via emergent SU(3) flux on the antiferromagnetic skyrmion lattice}},
  author = {Takeda, Hikaru and Kawano, Masataka and Tamura, Kyo and Akazawa, Masatoshi and Yan, Jian and Waki, Takeshi and Nakamura, Hiroyuki and Sato, Kazuki and Narumi, Yasuo and Hagiwara, Masayuki and Yamashita, Minoru and Hotta, Chisa},
  journal = {Nat. Commun.},
  volume = {15},
  pages = {566},
  year = {2024},
  doi = {10.1038/s41467-024-44793-3},
  url = {https://doi.org/10.1038/s41467-024-44793-3}
}

@article{kawano2025prb,
  title = {{Thermal Hall effect and gauge-field picture of magnons in antiferromagnetic skyrmion crystals}},
  author = {Kawano, Masataka},
  journal = {Phys. Rev. B},
  volume = {112},
  issue = {6},
  pages = {L060403},
  numpages = {7},
  year = {2025},
  month = {Aug},
  publisher = {American Physical Society},
  doi = {10.1103/vqny-81z1},
  url = {https://link.aps.org/doi/10.1103/vqny-81z1}
}

@article{kawano2025axv,
  title = {{Effective field theory and thermal Hall effect of magnons in square-lattice antiferromagnets}},
  author = {Kawano, Masataka},
  journal = {arXiv:2502.11924},
  year = {2025},
  url = {https://arxiv.org/abs/2502.11924}
}

@article{nakatsuji2015nature,
  title = {{Large anomalous Hall effect in a non-collinear antiferromagnet at room temperature}},
  author = {Nakatsuji, Satoru and Kiyohara, Naoki and Higo, Tomoya},
  journal = {Nature},
  volume = {527},
  pages = {212-215},
  year = {2015},
  doi = {10.1038/nature15723},
  url = {https://doi.org/10.1038/nature15723}
}

@article{nayak2016sciad,
  title = {{Large anomalous Hall effect driven by a nonvanishing Berry curvature in the noncolinear antiferromagnet Mn$_{3}$Ge}},
  author = {Ajaya K. Nayak and Julia Erika Fischer and Yan Sun and Binghai Yan and Julie Karel and Alexander C. Komarek and Chandra Shekhar and Nitesh Kumar and Walter Schnelle and J\"{u}rgen K\"{u}bler and Claudia Felser and Stuart S. P. Parkin},
  journal = {Sci. Adv.},
  volume = {2},
  pages = {1501870},
  year = {2016},
  doi = {10.1126/sciadv.1501870},
  url = {https://www.science.org/doi/abs/10.1126/sciadv.1501870}
}

@article{kiyohara2016prap,
  title = {{Giant Anomalous Hall Effect in the Chiral Antiferromagnet ${\mathrm{Mn}}_{3}\mathrm{Ge}$}},
  author = {Kiyohara, Naoki and Tomita, Takahiro and Nakatsuji, Satoru},
  journal = {Phys. Rev. Appl.},
  volume = {5},
  issue = {6},
  pages = {064009},
  numpages = {10},
  year = {2016},
  month = {Jun},
  publisher = {American Physical Society},
  doi = {10.1103/PhysRevApplied.5.064009},
  url = {https://link.aps.org/doi/10.1103/PhysRevApplied.5.064009}
}

@article{liu2017srep,
  title = {{Transition from Anomalous Hall Effect to Topological Hall Effect in Hexagonal Non-Collinear Magnet Mn$_{3}$Ga}},
  author = {Liu, Z. H. and Zhang, Y. J. and Liu, G. D. and Ding, B. and Liu, E. K. and Jafri, Hasnain Mehdi and Hou, Z. P. and Wang, W. H. and Ma, X. Q. and Wu, G. H.},
  journal = {Sci. Rep.},
  volume = {7},
  pages = {515},
  year = {2017},
  doi = {10.1038/s41598-017-00621-x},
  url = {https://doi.org/10.1038/s41598-017-00621-x}
}

@article{reil2002jac,
  title = {{Structural investigations of the compounds ASc2S4 (A=Mn, Fe, Cd)}},
  author = {S. Reil and H.-J. Stork and H. Haeuseler},
  journal = {J. Alloy. Compd.},
  volume = {334},
  pages = {92-96},
  year = {2002},
  doi = {https://doi.org/10.1016/S0925-8388(01)01772-8},
  url = {https://www.sciencedirect.com/science/article/pii/S0925838801017728}
}

@article{fritsch2004prl,
  title = {{Spin and Orbital Frustration in {M}n{S}c$_2${S}$_4$ and {F}e{S}c$_2${S}$_4$}},
  author = {Fritsch, V. and Hemberger, J. and B\"uttgen, N. and Scheidt, E.-W. and Krug von Nidda, H.-A. and Loidl, A. and Tsurkan, V.},
  journal = {Phys. Rev. Lett.},
  volume = {92},
  issue = {11},
  pages = {116401},
  numpages = {4},
  year = {2004},
  month = {Mar},
  publisher = {American Physical Society},
  doi = {10.1103/PhysRevLett.92.116401},
  url = {https://link.aps.org/doi/10.1103/PhysRevLett.92.116401}
}

@article{gao2017nphys,
  title = {{Spiral spin-liquid and the emergence of a vortex-like state in {M}n{S}c$_2${S}$_4$}},
  author = {Gao, S. and Zaharko, O. and Tsurkan, V. and Su, Y. and White, J.S. and Tucker, G.S. and Roessli, B. and Bourdarot, F. and Sibille, R. and Chernyshov, D. and Fennell, T. and Loidl, A. and D.C. and R\"uegg, C.},
  journal = {Nature Phys.},
  volume = {13},
  pages = {157},
  year = {2017},
  month = {Oct},
  publisher = {Nature publishing group},
  doi = {10.1038/NPHYS3914},
  url = {https://doi.org/10.1038/nphys3914}
}

@article{gao2020nature,
  title = {{Fractional antiferromagnetic skyrmion lattice induced by anisotropic couplings}},
  author = {Gao, S. and Rosales, H. D. and Albarracin, F. A .G. and Tsurkan, G. K. and Fennell, T. and Steffens, P. and Boehm, M. and Cermak, P. and Schneldewind, A. and Ressouche, E. and Cabra, D.C. and R\"uegg, C. and Zaharko, O.},
  journal = {Nature},
  volume = {586},
  pages = {37},
  year = {2020},
  month = {Oct},
  publisher = {Nature publishing group},
  doi = {10.1038/s41586-020-2716-8},
  url = {https://doi.org/10.1038/s41586-020-2716-8}
}

@article{mermin1976prl,
  title = {{Circulation and Angular Momentum in the $A$ Phase of Superfluid Helium-3}},
  author = {Mermin, N. D. and Ho, Tin-Lun},
  journal = {Phys. Rev. Lett.},
  volume = {36},
  issue = {11},
  pages = {594--597},
  numpages = {0},
  year = {1976},
  month = {Mar},
  publisher = {American Physical Society},
  doi = {10.1103/PhysRevLett.36.594},
  url = {https://link.aps.org/doi/10.1103/PhysRevLett.36.594}
}

@article{hua2014prl,
  title = {{Anomalous Hall Effect Arising from Noncollinear Antiferromagnetism}},
  author = {Chen, Hua and Niu, Qian and MacDonald, A. H.},
  journal = {Phys. Rev. Lett.},
  volume = {112},
  issue = {1},
  pages = {017205},
  numpages = {5},
  year = {2014},
  month = {Jan},
  publisher = {American Physical Society},
  doi = {10.1103/PhysRevLett.112.017205},
  url = {https://link.aps.org/doi/10.1103/PhysRevLett.112.017205}
}

@article{iguchi2018prb,
  title = {{Microwave nonreciprocity of magnon excitations in the noncentrosymmetric antiferromagnet ${\mathrm{Ba}}_{2}{\mathrm{MnGe}}_{2}{\mathrm{O}}_{7}$}},
  author = {Iguchi, Y. and Nii, Y. and Kawano, M. and Murakawa, H. and Hanasaki, N. and Onose, Y.},
  journal = {Phys. Rev. B},
  volume = {98},
  issue = {6},
  pages = {064416},
  numpages = {5},
  year = {2018},
  month = {Aug},
  publisher = {American Physical Society},
  doi = {10.1103/PhysRevB.98.064416},
  url = {https://link.aps.org/doi/10.1103/PhysRevB.98.064416}
}

@article{ding-zhang2025prl,
  title = {{Intrinsic Thermal Hall Effect in Mott Insulators}},
  author = {Ding, Jixun K. and Zhang, Emily Z. and Wang, Wen O. and Cookmeyer, Tessa and Moritz, Brian and Kim, Yong Baek and Devereaux, Thomas P.},
  journal = {Phys. Rev. Lett.},
  volume = {134},
  issue = {25},
  pages = {256501},
  numpages = {7},
  year = {2025},
  month = {Jun},
  publisher = {American Physical Society},
  doi = {10.1103/2729-nmyh},
  url = {https://link.aps.org/doi/10.1103/2729-nmyh}
}

@article{nagaosa2010rmp,
  title = {{Anomalous Hall effect}},
  author = {Nagaosa, Naoto and Sinova, Jairo and Onoda, Shigeki and MacDonald, A. H. and Ong, N. P.},
  journal = {Rev. Mod. Phys.},
  volume = {82},
  issue = {2},
  pages = {1539--1592},
  numpages = {0},
  year = {2010},
  month = {May},
  publisher = {American Physical Society},
  doi = {10.1103/RevModPhys.82.1539},
  url = {https://link.aps.org/doi/10.1103/RevModPhys.82.1539}
}

@article{chumak2015natphys,
  title = {{Magnon spintronics}},
  author = {Chumak, Andrii V and Vasyuchka, Vitaliy I and Serga, Alexander A and Hillebrands, Burkard},
  journal = {Nat. Phys.},
  volume = {11},
  number = {6},
  pages = {453--461},
  year = {2015},
  publisher = {Nature Publishing Group UK London},
  doi = {10.1038/nphys3347},
  url = {https://doi.org/10.1038/nphys3347}
}

@article{zhang2024pr,
title = {{Thermal Hall effects in quantum magnets}},
author = {Xiao-Tian Zhang and Yong Hao Gao and Gang Chen},
journal = {Phys. Rep.},
volume = {1070},
pages = {1-59},
year = {2024},
issn = {0370-1573},
doi = {https://doi.org/10.1016/j.physrep.2024.03.004},
url = {https://www.sciencedirect.com/science/article/pii/S0370157324001194}
}

@article{kane2005prl1,
  title = {{Quantum Spin Hall Effect in Graphene}},
  author = {Kane, C. L. and Mele, E. J.},
  journal = {Phys. Rev. Lett.},
  volume = {95},
  issue = {22},
  pages = {226801},
  numpages = {4},
  year = {2005},
  month = {Nov},
  publisher = {American Physical Society},
  doi = {10.1103/PhysRevLett.95.226801},
  url = {https://link.aps.org/doi/10.1103/PhysRevLett.95.226801}
}

@article{kane2005prl2,
  title = {{${Z}_{2}$ Topological Order and the Quantum Spin Hall Effect}},
  author = {Kane, C. L. and Mele, E. J.},
  journal = {Phys. Rev. Lett.},
  volume = {95},
  issue = {14},
  pages = {146802},
  numpages = {4},
  year = {2005},
  month = {Sep},
  publisher = {American Physical Society},
  doi = {10.1103/PhysRevLett.95.146802},
  url = {https://link.aps.org/doi/10.1103/PhysRevLett.95.146802}
}

@article{hayami2020prb,
  title = {{Bottom-up design of spin-split and reshaped electronic band structures in antiferromagnets without spin-orbit coupling: Procedure on the basis of augmented multipoles}},
  author = {Hayami, Satoru and Yanagi, Yuki and Kusunose, Hiroaki},
  journal = {Phys. Rev. B},
  volume = {102},
  issue = {14},
  pages = {144441},
  numpages = {24},
  year = {2020},
  month = {Oct},
  publisher = {American Physical Society},
  doi = {10.1103/PhysRevB.102.144441},
  url = {https://link.aps.org/doi/10.1103/PhysRevB.102.144441}
}

@article{smejkal2022prx1,
  title = {{Emerging Research Landscape of Altermagnetism}},
  author = {\ifmmode \check{S}\else \v{S}\fi{}mejkal, Libor and Sinova, Jairo and Jungwirth, Tomas},
  journal = {Phys. Rev. X},
  volume = {12},
  issue = {4},
  pages = {040501},
  numpages = {27},
  year = {2022},
  month = {Dec},
  publisher = {American Physical Society},
  doi = {10.1103/PhysRevX.12.040501},
  url = {https://link.aps.org/doi/10.1103/PhysRevX.12.040501}
}

@article{smejkal2022prx2,
  title = {{Beyond Conventional Ferromagnetism and Antiferromagnetism: A Phase with Nonrelativistic Spin and Crystal Rotation Symmetry}},
  author = {\ifmmode \check{S}\else \v{S}\fi{}mejkal, Libor and Sinova, Jairo and Jungwirth, Tomas},
  journal = {Phys. Rev. X},
  volume = {12},
  issue = {3},
  pages = {031042},
  numpages = {16},
  year = {2022},
  month = {Sep},
  publisher = {American Physical Society},
  doi = {10.1103/PhysRevX.12.031042},
  url = {https://link.aps.org/doi/10.1103/PhysRevX.12.031042}
}

@article{zhitomirsky2013rmp,
  title = {{Colloquium: Spontaneous magnon decays}},
  author = {Zhitomirsky, M. E. and Chernyshev, A. L.},
  journal = {Rev. Mod. Phys.},
  volume = {85},
  issue = {1},
  pages = {219--242},
  numpages = {0},
  year = {2013},
  month = {Jan},
  publisher = {American Physical Society},
  doi = {10.1103/RevModPhys.85.219},
  url = {https://link.aps.org/doi/10.1103/RevModPhys.85.219}
}

@article{grissonnanche2019nature,
  title={{Giant thermal Hall conductivity in the pseudogap phase of cuprate superconductors}},
  author={Grissonnanche, Ga{\"e}l and Legros, Ana{\"e}lle and Badoux, Sven and Lefran{\c{c}}ois, Etienne and Zatko, Victor and Lizaire, Maude and Lalibert{\'e}, Francis and Gourgout, Adrien and Zhou, J-S and Pyon, Sunseng and others},
  journal={Nature},
  volume={571},
  number={7765},
  pages={376--380},
  year={2019},
  publisher={Nature Publishing Group UK London},
  url={https://www.nature.com/articles/s41586-019-1375-0}
}

@article{boulanger2020ncom,
  title={{Thermal Hall conductivity in the cuprate Mott insulators Nd2CuO4 and Sr2CuO2Cl2}},
  author={Boulanger, Marie-Eve and Grissonnanche, Ga{\"e}l and Badoux, Sven and Allaire, Andr{\'e}anne and Lefran{\c{c}}ois, {\'E}tienne and Legros, Ana{\"e}lle and Gourgout, Adrien and Dion, Maxime and Wang, CH and Chen, XH and others},
  journal={Nature communications},
  volume={11},
  number={1},
  pages={5325},
  year={2020},
  publisher={Nature Publishing Group UK London},
  url={https://www.nature.com/articles/s41467-020-18881-z#citeas}
}

@article{chen2024ncom,
  title={{Planar thermal Hall effect from phonons in a Kitaev candidate material}},
  author={Chen, Lu and Lefran{\c{c}}ois, {\'E}tienne and Vallipuram, Ashvini and Barth{\'e}lemy, Quentin and Ataei, Amirreza and Yao, Weiliang and Li, Yuan and Taillefer, Louis},
  journal={Nat. Commun.},
  volume={15},
  number={1},
  pages={3513},
  year={2024},
  publisher={Nature Publishing Group UK London},
  url={https://www.nature.com/articles/s41467-024-47858-5}
}

@article{strohm2005prl,
  title = {{Phenomenological Evidence for the Phonon Hall Effect}},
  author = {Strohm, C. and Rikken, G. L. J. A. and Wyder, P.},
  journal = {Phys. Rev. Lett.},
  volume = {95},
  issue = {15},
  pages = {155901},
  numpages = {4},
  year = {2005},
  month = {Oct},
  publisher = {American Physical Society},
  doi = {10.1103/PhysRevLett.95.155901},
  url = {https://link.aps.org/doi/10.1103/PhysRevLett.95.155901}
}

@article{qin2012prb,
  title = {{Berry curvature and the phonon Hall effect}},
  author = {Qin, Tao and Zhou, Jianhui and Shi, Junren},
  journal = {Phys. Rev. B},
  volume = {86},
  issue = {10},
  pages = {104305},
  numpages = {9},
  year = {2012},
  month = {Sep},
  publisher = {American Physical Society},
  doi = {10.1103/PhysRevB.86.104305},
  url = {https://link.aps.org/doi/10.1103/PhysRevB.86.104305}
}

@article{dhakal2024axv,
  title={{Theory of Intrinsic Phonon Thermal Hall Effect in $$\backslash$alpha $-RuCl $ \_3$}},
  author={Dhakal, Ramesh and Kaib, David AS and Choi, Kate and Biswas, Sananda and Valenti, Roser and Winter, Stephen M},
  journal={arXiv preprint arXiv:2407.00660},
  year={2024},
  url={https://arxiv.org/abs/2407.00660}
}

@article{chen2024prx,
  title = {{Planar Thermal Hall Effect from Phonons in Cuprates}},
  author = {Chen, Lu and Le Roux, L\'ena and Grissonnanche, Ga\"el and Boulanger, Marie-Eve and Th\'eriault, Steven and Liang, Ruixing and Bonn, D. A. and Hardy, W. N. and Pyon, S. and Takayama, T. and Takagi, H. and Xu, Ke-Jun and Shen, Zhi-Xun and Taillefer, Louis},
  journal = {Phys. Rev. X},
  volume = {14},
  issue = {4},
  pages = {041011},
  numpages = {9},
  year = {2024},
  month = {Oct},
  publisher = {American Physical Society},
  doi = {10.1103/PhysRevX.14.041011},
  url = {https://link.aps.org/doi/10.1103/PhysRevX.14.041011}
}

@article{frohlich1993rmp,
  title = {{Gauge invariance and current algebra in nonrelativistic many-body theory}},
  author = {Fr\"ohlich, J\"urg and Studer, Urban M.},
  journal = {Rev. Mod. Phys.},
  volume = {65},
  pages = {733--802},
  year = {1993},
  url = {https://link.aps.org/doi/10.1103/RevModPhys.65.733}
}

@misc{comment-triangular, note={{Although a thermal Hall signal has been reported in the noncoplanar antiferromagnet in the distorted triangular lattice YMnO$_{3}$, it is attributed to topological spin fluctuations and phonon-related effects, rather than to a magnon thermal Hall effect.}},
}

@misc{comment-su2, note={{Precisely speaking, for the SU(2) gauge field, an effective Zeeman term is also necessary to circumvent the no-go rule.}}}

@article{debnath2025prb1,
  title = {{Magnons on a dice lattice: Topological features and transport properties}},
  author = {Debnath, Shreya and Basu, Saurabh},
  journal = {Phys. Rev. B},
  volume = {111},
  issue = {15},
  pages = {155418},
  numpages = {20},
  year = {2025},
  month = {Apr},
  publisher = {American Physical Society},
  doi = {10.1103/PhysRevB.111.155418},
  url = {https://link.aps.org/doi/10.1103/PhysRevB.111.155418}
}

@article{zhang2019prl,
  title = {{Thermal Hall Effect Induced by Magnon-Phonon Interactions}},
  author = {Zhang, Xiaoou and Zhang, Yinhan and Okamoto, Satoshi and Xiao, Di},
  journal = {Phys. Rev. Lett.},
  volume = {123},
  issue = {16},
  pages = {167202},
  numpages = {6},
  year = {2019},
  month = {Oct},
  publisher = {American Physical Society},
  doi = {10.1103/PhysRevLett.123.167202},
  url = {https://link.aps.org/doi/10.1103/PhysRevLett.123.167202}
}

@article{park2020nl,
  title = {{Thermal Hall Effect, Spin Nernst Effect, and Spin Density Induced by a Thermal Gradient in Collinear Ferrimagnets from Magnon-Phonon Interaction}},
  author = {Park, Sungjoon and Nagaosa, Naoto and Yang, Bohm-Jung},
  journal = {Nano Lett.},
  volume = {20},
  pages = {2741},
  year = {2020},
  doi = {10.1021/acs.nanolett.0c00363},
  url = {https://doi.org/10.1021/acs.nanolett.0c00363}
}

@article{akazawa2020prx,
  title = {{Thermal Hall Effects of Spins and Phonons in Kagome Antiferromagnet Cd-Kapellasite}},
  author = {Akazawa, Masatoshi and Shimozawa, Masaaki and Kittaka, Shunichiro and Sakakibara, Toshiro and Okuma, Ryutaro and Hiroi, Zenji and Lee, Hyun-Yong and Kawashima, Naoki and Han, Jung Hoon and Yamashita, Minoru},
  journal = {Phys. Rev. X},
  volume = {10},
  issue = {4},
  pages = {041059},
  numpages = {14},
  year = {2020},
  month = {Dec},
  publisher = {American Physical Society},
  doi = {10.1103/PhysRevX.10.041059},
  url = {https://link.aps.org/doi/10.1103/PhysRevX.10.041059}
}

@article{li2023prb,
  title = {{Magnon-polaron driven thermal Hall effect in a Heisenberg-Kitaev antiferromagnet}},
  author = {Li, N. and Neumann, R. R. and Guang, S. K. and Huang, Q. and Liu, J. and Xia, K. and Yue, X. Y. and Sun, Y. and Wang, Y. Y. and Li, Q. J. and Jiang, Y. and Fang, J. and Jiang, Z. and Zhao, X. and Mook, A. and Henk, J. and Mertig, I. and Zhou, H. D. and Sun, X. F.},
  journal = {Phys. Rev. B},
  volume = {108},
  issue = {14},
  pages = {L140402},
  numpages = {7},
  year = {2023},
  month = {Oct},
  publisher = {American Physical Society},
  doi = {10.1103/PhysRevB.108.L140402},
  url = {https://link.aps.org/doi/10.1103/PhysRevB.108.L140402}
}

@article{debnath2025prb2,
  title = {{Topological characterization of magnon-polaron bands and thermal Hall conductivity in a frustrated kagome antiferromagnet}},
  author = {Debnath, Shreya and Bhattacharyya, Kuntal and Basu, Saurabh},
  journal = {Phys. Rev. B},
  volume = {112},
  issue = {12},
  pages = {125404},
  numpages = {19},
  year = {2025},
  month = {Sep},
  publisher = {American Physical Society},
  doi = {10.1103/w9rc-m6qf},
  url = {https://link.aps.org/doi/10.1103/w9rc-m6qf}
}

@article{ideue2017nmat,
  title = {{Giant thermal Hall effect in multiferroics}},
  author = {Ideue, T. and Kurumaji, T. and Ishiwata, S. and Tokura, Y.},
  journal = {Nat. Mat.},
  volume = {16},
  pages = {797},
  year = {2017},
  doi = {10.1038/nmat4905},
  url = {https://doi.org/10.1038/nmat4905}
}

\end{document}